\documentclass{PoS}
\usepackage{amsmath}
\usepackage{amssymb}
\usepackage{graphicx}
\usepackage{epsfig}
\usepackage{color}
\usepackage{latexsym}
\usepackage{amsmath}
\usepackage{amssymb}
\usepackage{dsfont}
\usepackage{verbatim}  

\title{Hadron Spectroscopy}

\ShortTitle{Hadron spectroscopy}

\author{\speaker{Sasa Prelovsek}\\
       Faculty of Mathematics and Physics, University of Ljubljana, Jadranska 19, Ljubljana, Slovenia\\
       and Jozef Stefan Institute, Jamova 39, Ljubljana, Slovenia\\
        E-mail: \email{sasa.prelovsek@ijs.si}}

\abstract{ Recent results on the hadron spectroscopy from lattice QCD are reviewed with emphasis on the meson sector and in particular on quarkonium-like $XYZ$ states.  I report on the first rigorous treatment of the near-threshold states $X(3872)$ and  $D_s^0(2317)$, and the   lattice searches for $Z_c^+(3900)$, $X(4140)$ and  double-charm tetraquark states.  Meson resonances in light, strange and charm sector are reviewed, where  the resonances masses as well as the  strong decay widths are reported.  The first lattice QCD simulation of two coupled-channels is discussed.   }

\FullConference{The 32nd International Symposium on Lattice Field Theory,\\
		23-28 June, 2014\\
		Columbia University New York, NY}

\begin{document}

\section{Introduction}

I was   requested to review the hadron spectroscopy focusing on mesons with particular emphasis on quarkonium-like  states. Indeed, there is a compelling motivation at present to establish whether such hadrons arise from first-principle QCD or not. Several experiments have, for example, found resonant structures in the $J/\psi\,  \pi^+$ invariant mass, indicating the existence of hadrons with exotic flavor composition $\bar cc\bar d u$ (see \cite{Brambilla:2014jmp} for review). 

Therefore I will concentrate on the recent lattice results on the meson spectroscopy and separately discuss hadrons well-below strong decay threshold, those near threshold and resonances above threshold.  Simulations of the exotic flavor channels will be reported, where I will provide recent results and also refer to some of the older simulations.

 \section{ The discrete spectrum from lattice and information encoded in it}

The physics information on a hadron (below, near or above threshold) is commonly extracted from the discrete energy spectrum in lattice QCD. 
The physical system for given quantum numbers is created from the vacuum $|\Omega\rangle$ using interpolator ${\cal O}_j^\dagger$  at time $t\!=\!0$ and the system propagates for time $t$ before being annihilated by ${\cal O}_i$.    To study a meson state with given $J^P$ one  can, for example,  use ${\cal O}\simeq \bar q	\Gamma q,~$ $(\bar q \Gamma_1  q)_{\vec p_1}(\bar q \Gamma_2  q)_{\vec p_2},~$$[\bar q \Gamma_1 \bar q][q\Gamma_2 q]$ with desired quantum numbers.  After the spectral decomposition the correlators are expressed in terms of the energies  $E_n$ of eigenstates $|n\rangle$ and their overlaps $Z_j^n$
\begin{equation}
\label{C}
C_{ij}(t)= \langle \Omega|{\cal O}_i (t) {\cal O}_j^\dagger (0)|\Omega \rangle=\sum_{n}Z_i^nZ_j^{n*}~e^{-E_n t}~,\qquad Z_i^n\equiv \langle \Omega|{\cal O}_i|n\rangle~.
\end{equation}
The most widely used method to extract  $E_n$ and  $Z_i^n$ from the correlation matrix $C_{ij}(t)$ is the generalized
eigenvalue method  $C(t)u^{n}(t)=\lambda^{n}(t)C(t_0)u^{n}(t)$ \cite{Michael:1985ne,Blossier:2009kd}.
The energies $E_n$ are obtained from the  exponential
behavior of the eigenvalues $\lambda^{n}(t)\propto  e^{-E_n t}$ at large $t$.   

 All physical eigenstates with given quantum numbers appear as energy levels in principle. These can be  "one-hadron" states, "two-hadron" states  and the multi-hadron states.  In reality the eigenstates are mixtures of these Fock components. Three- and more-hadron states have not been taken into account in the actual  simulations for the  spectroscopy of the hadronic  resonances yet\footnote{Multi-hadron states have been, for example, simulated to determine the binding energies of the nuclei.}. The major step during the past few years came from treating two-hadron states rigorously. 
 These  have a discrete spectrum due to the periodic boundary condition on finite lattices.  If the two hadrons do not interact, then the momenta of each hadron is $\vec{p}= \!\tfrac{2\pi}{L}\vec{N}$ with $\vec{N}\in {N}^3$, and the non-interacting energies of $H_1(\vec p)H_2(-\vec p)$    are $E^{n.i.}=E_1(p)+E_2(p)$ with $E_{1,2}(p)=(m_{1,2}^2+p^2)^{1/2}$.  The energies $E_n$ extracted from the lattice  are slightly shifted in presence of the interaction and the shift  provides rigorous information on the scattering matrix, as discussed below. 
In experiment, two-hadron  states correspond to the two-hadron decay products with a continuous energy spectrum. 

Particular care has to be taken concerning the discretization errors related to the heavy quarks. It is comforting to see that the complementary methods lead to compatible results in the continuum limit. For hadrons containing the charm quarks it is common to compare $m-m_{reference}$ between lattice and experiment, where the leading discretization errors related to $m_c$ cancel. 

The evaluation of the correlation matrices including $\bar qq$ as well as $\bar qq\bar qq$ interpolating fields entails various Wick contractions, which often require quark propagators from any source to any sink location. These are not rendered by the conventional method, and  all-to-all methods such as distillation \cite{Peardon:2009gh}, stochastic distillation \cite{Morningstar:2011ka} or noise reduction technique \cite{Bali:2009hu}  are applied. These are  based on a particular (separable) form for quark smearing  and are particularly useful for hadron spectroscopy since the spectrum $E_n$ is independent of the shape for quark smearing. 

All simulations of charmonia and other hidden charm  channels reported here omit the charm-quark annihilation contribution, while possible Wick contractions with $u/d/s$ annihilation are taken into account.  The  OZI-suppressed charm-annihilation  represents mixing  with channels that contain no charm quarks, and these decay channels  are indeed suppressed or unobserved in experiment.      The rigorous treatment of  this Wick contraction presents an unsolved problem on the lattice due to the mixing with a number of light hadron channels and the noise in the charm disconnected diagrams.  

  \section{Information encoded in the finite-volume spectrum}
 
  Most of the physics information on the hadron masses and strong interactions between them has been obtained from the  finite volume spectrum $E_n$. The mass of a hadron well below strong decay threshold is simply $m=E\vert_{\vec{p}=0}$. In the energy region near or above threshold the masses of bound-states and resonances have to be inferred from the infinite-volume scattering matrix of the one-channel (elastic) or multiple-channel (inelastic) scattering - I will refer to it as the rigorous treatment.  The bound states correspond to poles of the scattering matrix  at $p^2<0$  below threshold.    The resonances masses and widths are derived from the Breit-Wigner type fits of the scattering matrix, cross-section or phase shift.  
    
  Three approaches for extracting the  infinite-volume scattering matrix from the finite volume $E_n$  are employed  in the lattice studies reported here.  The methods were reviewed  in more detail during  plenary talks \cite{Yamazaki:lat14,Briceno:lat14} at this meeting:
\begin{enumerate}
\item  The most widely used approach  is based on L\"uscher's seminal work \cite{Luscher:1985dn,Luscher:1986pf,Luscher:1990ux,Luscher:1991cf} and its generalizations.  In the case of elastic scattering between two hadrons with zero-total momentum, the energy $E_n=(m_1^2+p^2)^{1/2}+(m_2^2+p^2)^{1/2}$ renders momenta $p$  of each meson in the region outside the interaction. The infinite volume phase shift at that energy is given by the L\"uscher's relation $\delta_l(p)=\mathrm{atan}[\sqrt{\pi} p L/2\,Z_{00}(1;(pL/2\pi)^2)]$  if the partial wave $l$ dominates the scattering \cite{Luscher:1991cf}. 
This is a favorable case where one equation determines one unknown $\delta_l(E_n)$ for each energy level $E_n$.    

The generalizations of this relation to multiple partial waves, non-zero total momenta, twisted boundary conditions\footnote{Partial twisting only in the valence sector represents an approximation if L\"uscher's relations based on the full twisting are employed.}, coupled-channel scattering and three-particle systems have also been derived in a series of papers recently (for reviews see \cite{Yamazaki:lat14,Briceno:lat14}).  For each energy level $E_n$ this generally leads to one (determinant) equation with several unknown $\delta_l^{a}(E_n^{cm})$ and the rigorous extraction becomes much more challenging. In this case the analysis may relay on certain parametrizations of the scattering matrix as a function of $E^{cm}$, which may render the otherwise unsolvable problem tractable. It is encouraging that the Hadron Spectrum Collaboration presented  the first simulation of two-coupled channel  system $K\pi-K\eta$ and extracted the poles corresponding to strange mesons relying on the parametrization of the scattering matrix \cite{Dudek:2014qha}.  
\item The finite volume Hamiltonian Effective Field Theory (EFT) represents an analogous approach. The parameters of this theory are extracted by fitting the analytic expressions for the eigenvalues of the finite-volume EFT using  the lattice spectrum $E_n$ \cite{Hall:2013qba,Leinweber:lat14}. 
\item HALQCD approach starts with the lattice determination of the Nambu-Bethe-Salpeter wave function for two hadrons  as a function of their separation, and extracts the potential $V(r)$ between two hadrons from that \cite{HALQCD:2012aa}. The phase shifts for scattering are obtained by solving the  Schr\"odinger equation for given $V(r)$.  Note that the potential is not a physics observable, while the resulting phase shift is.   It  was shown that HALQCD approach gives results in agreement with 
L\"uscher's approach for the $\pi\pi$ scattering with $I=2$ \cite{Charron:2013paa}, while systematic comparisons for other channels have not been done yet. There  are ongoing 
discussions as to whether this approach is as rigorous as the L\"uscher-type approach.  The method has recently been generalized to the case of multiple-channel scattering \cite{Aoki:2012bb} and preliminary results for the coupled channels $\Lambda\Lambda- N\Xi-\Sigma\Sigma$  \cite{Sasaki:2012ju} are particularly valuable as treatment of such a problem using the L\"uscher-type approach is very challenging. 
\end{enumerate} 

The resulting phase shifts enable the determination of the masses for  resonances and bound states. The simplest example is a one-channel elastic scattering  in partial wave $l$, where   the scattering matrix  is parametrized in terms of the phase shift $\delta_l(p)$ and satisfies unitarity $SS^\dagger=1$
\begin{equation}
\label{T}
S(p)=e^{2i\delta_l(p)}\ , \quad S(p)=1+2iT(p)\ , \quad T(p)=\frac{1}{\cot(\delta_l(p))-i}
\end{equation} 

\begin{itemize}
\item  In the vicinity of a {\it hadronic resonance} with mass $m_R$ and width $\Gamma$, the  cross section $\sigma \propto |T(p)|^2$ has a resonant shape with  $\delta(s=m_R^2)=\tfrac{\pi}{2}$
\begin{equation}
\label{R}
T(p)=\frac{-\sqrt{s}~ \Gamma(p)}{s-m_R^2+i \sqrt{s}\,\Gamma(p)}=\frac{1}{\cot\delta(p)-i}\ , \quad \Gamma(p)=g^2\,\frac{p^{2l+1}}{s}\ , \quad \frac{p^{2l+1}}{\sqrt{s}}\cot\delta(p)=\frac{1}{g^2}(m_R^2-s)\ . 
\end{equation}
This is  one possible  Breit-Wigner parametrization which are all equivalent for small $\Gamma$. The width $\Gamma$ is parametrized in terms of the phase  space and the coupling  $g$ (\ref{R}).  The fit of $\delta_l(p)$  renders $m_R$ and $g$ or $\Gamma$. It is customary to compare $g$ rather than $\Gamma$ to experiment, since $\Gamma$ depends on the masses of the scattering particles (for example $m_\pi$) via the phase space. 
\item The {\it bound state (B)} in $H_1 H_2$ scattering is realized when $T(p)$ has a pole at $p_B^2<0$  
\begin{equation}
\label{B}
T=  \frac{1}{\cot(\delta_l(p_B))-i}=\infty\ , \ \  \cot(\delta(p_B))=i\ ,\ \  m_B=E_{H_1}(p_B)+E_{H_2}(p_B)\ ,\ p_B= i|p_B|~.
\end{equation} 
 The  location of an s-wave shallow bound state can be obtained by parametrizing $\delta_0$ near threshold using the effective range expansion and finding $p_B$ which satisfies $\cot(\delta(p_B))=i$
 \begin{equation}
 \label{ER}
 p\, \cot(\delta_0(p))=\frac{1}{a_0}+\frac{1}{2}\,r_0\, p^2\ , \quad - |p_B| = \frac{1}{a_0}-\frac{1}{2}\,r_0\, |p_B|^2~.
\end{equation}
The large negative scattering length $a_0$ indicates the presence of a shallow bound state \cite{Sasaki:2006jn}. 
\end{itemize}

So far I discussed physics information that is obtained from the finite volume spectrum $E_n$. 
The overlaps $Z_i^n=\langle \Omega|{\cal O}_i|n\rangle$  provide a wealth of information about the composition of each lattice eigenstate $|n\rangle$. They have been  used so far  mostly as a qualitative guidance on the importance of various Fock components.  It remains an open question how to use this rich source of information to rigorously extract properties of the physical states, especially in the case of  smeared quarks. Analytic considerations in this direction may prove fruitful.

 \section{Hadrons well below threshold}
 
 Well below strong decay threshold there are no multi-hadron states, and the mass of a single hadron is extracted from $m=E\vert_{\vec{p}=0}$ extrapolated to $L\to \infty$, $a\to 0$ and $m_q\to m_q^{phys}$.  Many precision results have been available for a number of years.\\
 
\underline{${\mathbf{\bar cc}}$}:   The continuum and chiral extrapolations of   low-lying charmonia were, for example, addressed by Fermilab/MILC \cite{Mohler:lat14} and HPQCD/MILC \cite{Galloway:lat14} collaborations at this meeting. The resulting splittings between the ground-state masses in different channels as well as the spin-averaged masses of $2S$ and $1S$ charmonia  are in  good agreement with experiment after extrapolations.  \\ 

\underline{${\mathbf{\eta,\ \eta^\prime}}$}:  These isosinglet pseudoscalar mesons can strongly decay only to the three-meson states, therefore they are very narrow and can be treated using standard technique to a good approximation.  Their masses as well as the flavor mixing angle were determined as a function of $m_\pi$ by ETMC collaboration \cite{Michael:2013gka}, recovering experimental values in the chiral limit.  
 
\begin{figure}[htb] 
\begin{center}
\includegraphics*[width=0.90\textwidth,clip]{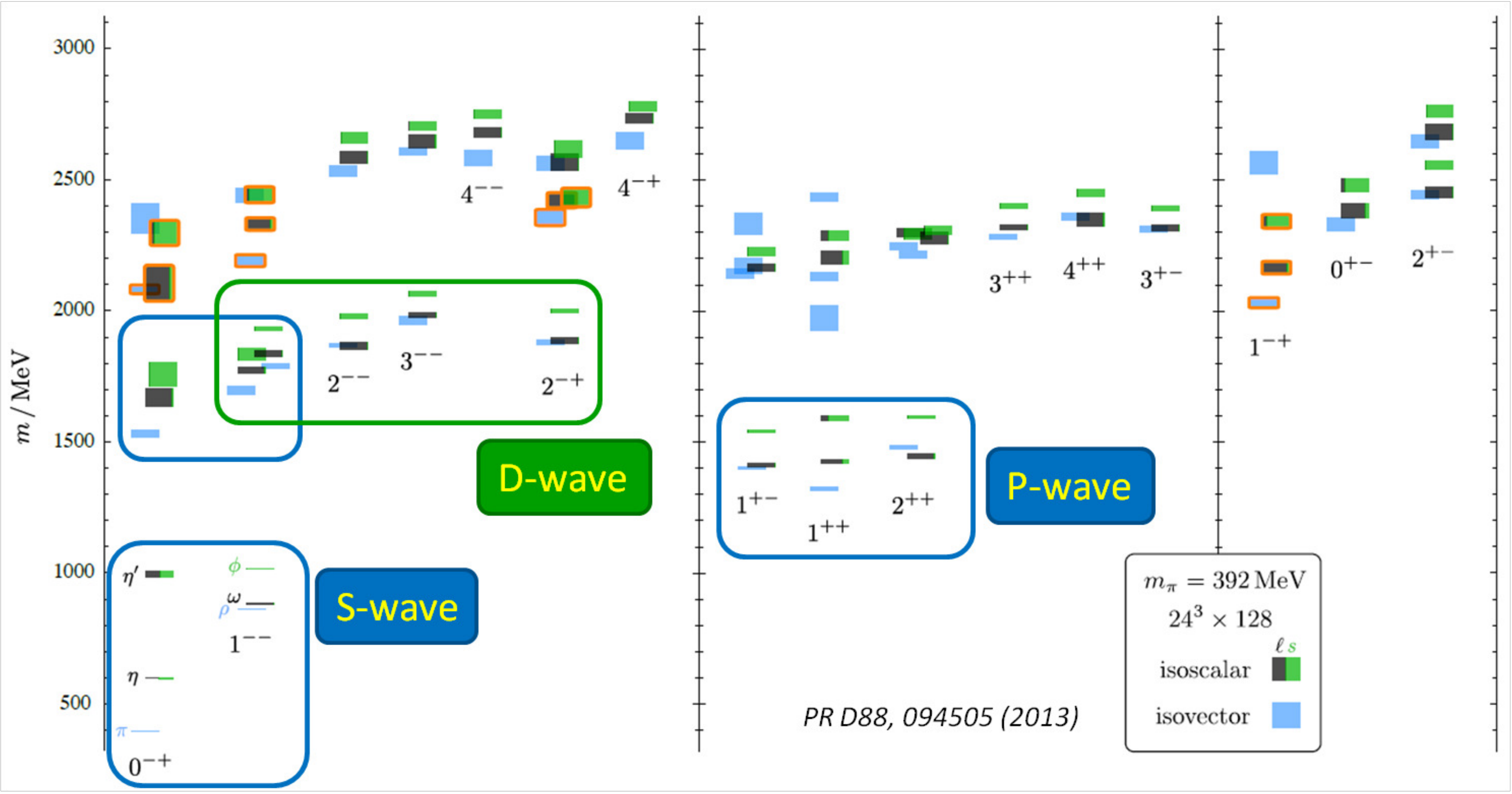} %
\caption{\label{fig:hsc_isoscalars}  The isoscalar meson spectrum:  black and green indicate percentage of  $|l\rangle=(\bar uu+\bar dd)/\sqrt{2}$ and $|s\rangle=\bar ss$ for each state, respectively \cite{Dudek:2013yja}. The blue levels show the masses of isovector mesons on the same ensemble.  }
\end{center}
\end{figure}

\section{Excited mesons within a single-hadron approach}

A great majority of hadrons lie near or above strong decay threshold. Yet most of them have been treated until recently  based on a single-hadron approach. This entails (i) using only ${\cal O}\simeq \bar qq$ for mesons and  ${\cal O}\simeq qqq$ for baryons, (ii) assuming that all energy levels correspond to ``one-particle'' states  and (iii) that the mass of the  excited resonance  equals $m\!=\!E$. These are strong assumptions for the resonances, which are not asymptotic states.  The approach also ignores the effect of the threshold on near-threshold states.

The most extensive light isoscalar  \cite{Dudek:2013yja}, $D$, $D_s$ \cite{Moir:2013ub} and $\bar cc$ \cite{Liu:2012ze} spectra were extracted  by the Hadron Spectrum Collaboration (HSC) on $N_f\!=\!2\!+\!1$ anisotropic configurations with $m_\pi\simeq 400~$MeV and two different $L\simeq 2.9~$fm,~$1.9~$fm. Their charmonium spectrum presents a valuable reference point for charmonium(like) states discussed later.  The spectrum of isoscalar mesons as well as their light and strange content  is shown in Fig. \ref{fig:hsc_isoscalars}.   An impressive number of excited states is extracted   with a good accuracy in spite of the disconnected contribution for isoscalars.  States  are identified with members of $\bar qq$ multiplets $nS$, $nP$ and $nD$ based on the overlaps $\langle {\cal O}_i|n\rangle$, where interpolators are chosen to resemble multiplet members.  
The remaining states   (indicated by orange) show particularly strong overlap with  ${\cal O}\simeq \bar q F_{\mu\nu}q$  and are identified as hybrids.

\section{Near-threshold  hadrons (beyond  single-hadron approach)}

 Most of the exciting states found by experiments are located near thresholds, for example $X(3872)$, $Z_c^+(3900)$, $Z_b^+(10610)$, $Z_b^+(10650)$,  $D_s^0(2317)$ and $\Lambda(1405)$. The quarkonium-like states near threshold and above threshold are listed in Tables 10 and 12 of a valuable  review  by Brambilla et al. \cite{Brambilla:2014jmp}. The lattice community is faced with an important challenge to establish  whether these states arise from the first-principle QCD or not, and what is their nature.
 
 Indeed, most of the effort in the hadron spectroscopy during past few years went in going beyond the single-hadron approach and taking into account two-hadron states rigorously. The majority of the studies focus on the (elastic) energy region near threshold, where the methods may be tractable at present, but one can not expect spectra of highly excited multiplets from the rigorous approach soon. \\

  \begin{figure}[htb] 
 \begin{center}
\includegraphics*[width=0.50\textwidth,clip]{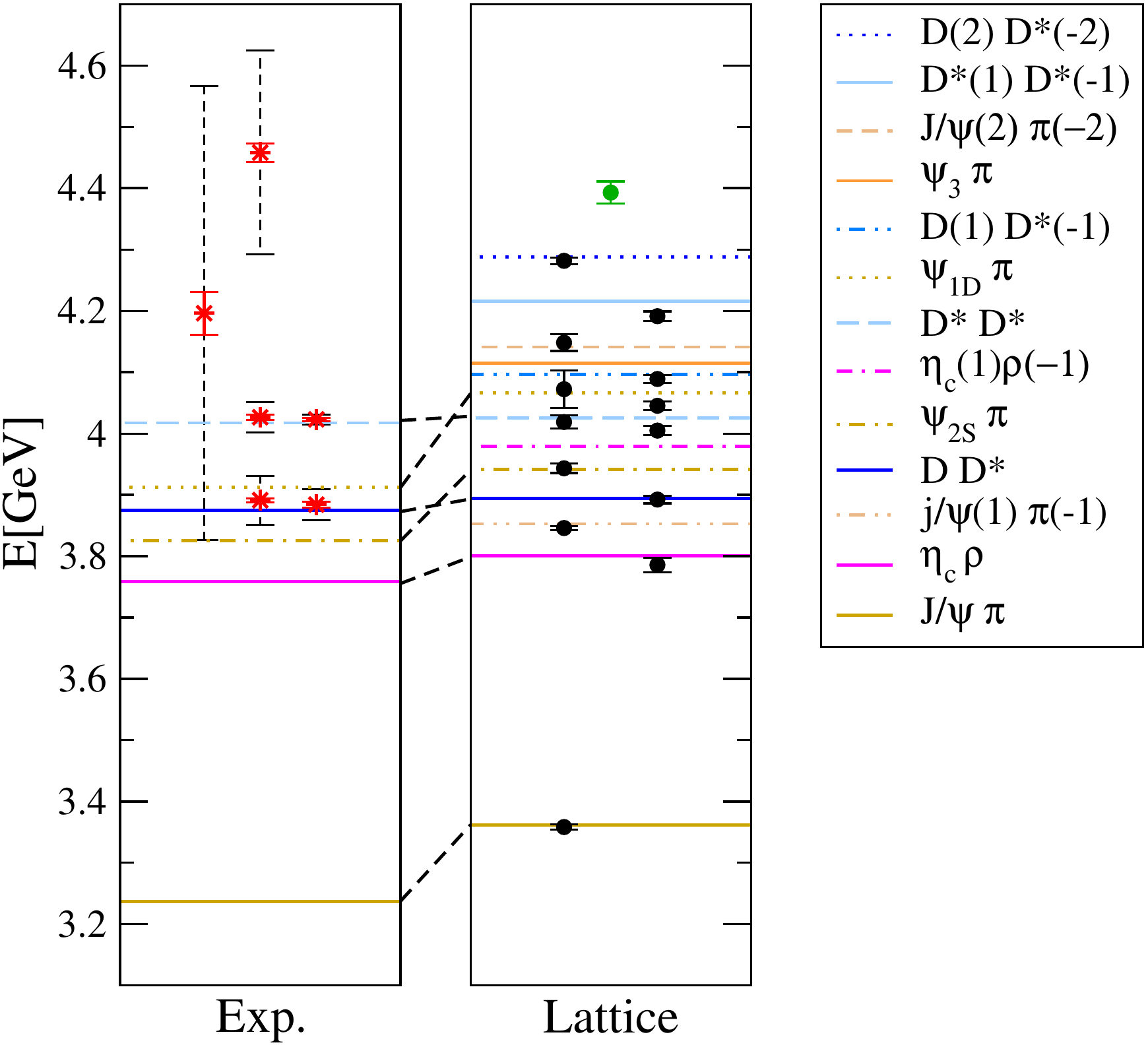} 
\caption{\label{fig:Zc_prelovsek}  Left: The masses of the experimental $Z_c^+\simeq \bar cc\bar d u$ candidates \cite{Brambilla:2014jmp} and the experimental thresholds.  Right:  The points represent the  lattice spectrum $E_n$ in the   $I^G(J^{PC})=1^+(1^{+-})$ channel  \cite{Prelovsek:2014v2}. Lines represent the energies of the non-interacting  two-meson states on this lattice. The states indicated by black circles are identified as two-meson states.   }
\end{center}
\end{figure}

\begin{figure}[htb] 
 \begin{center}
\includegraphics*[width=0.55\textwidth,clip]{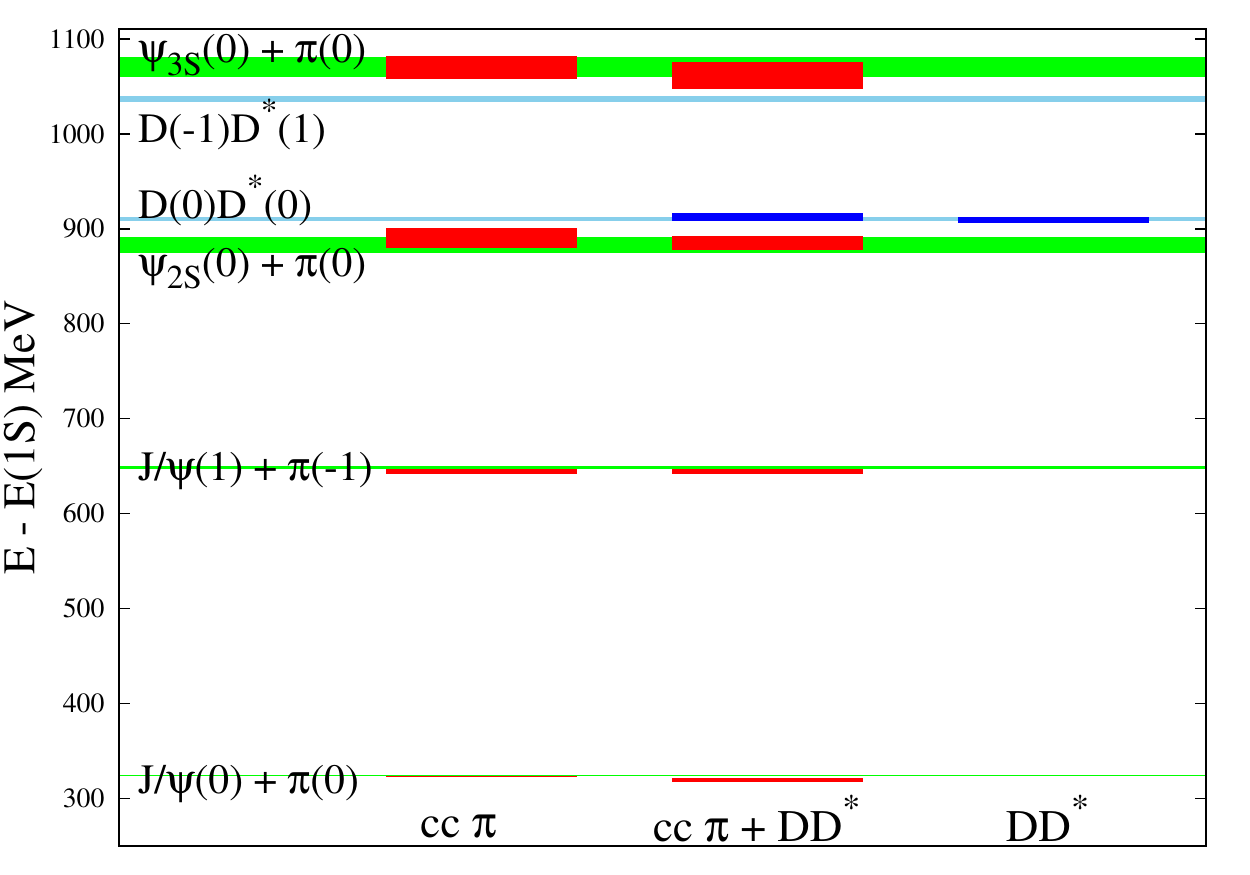}
\includegraphics*[width=0.55\textwidth,clip]{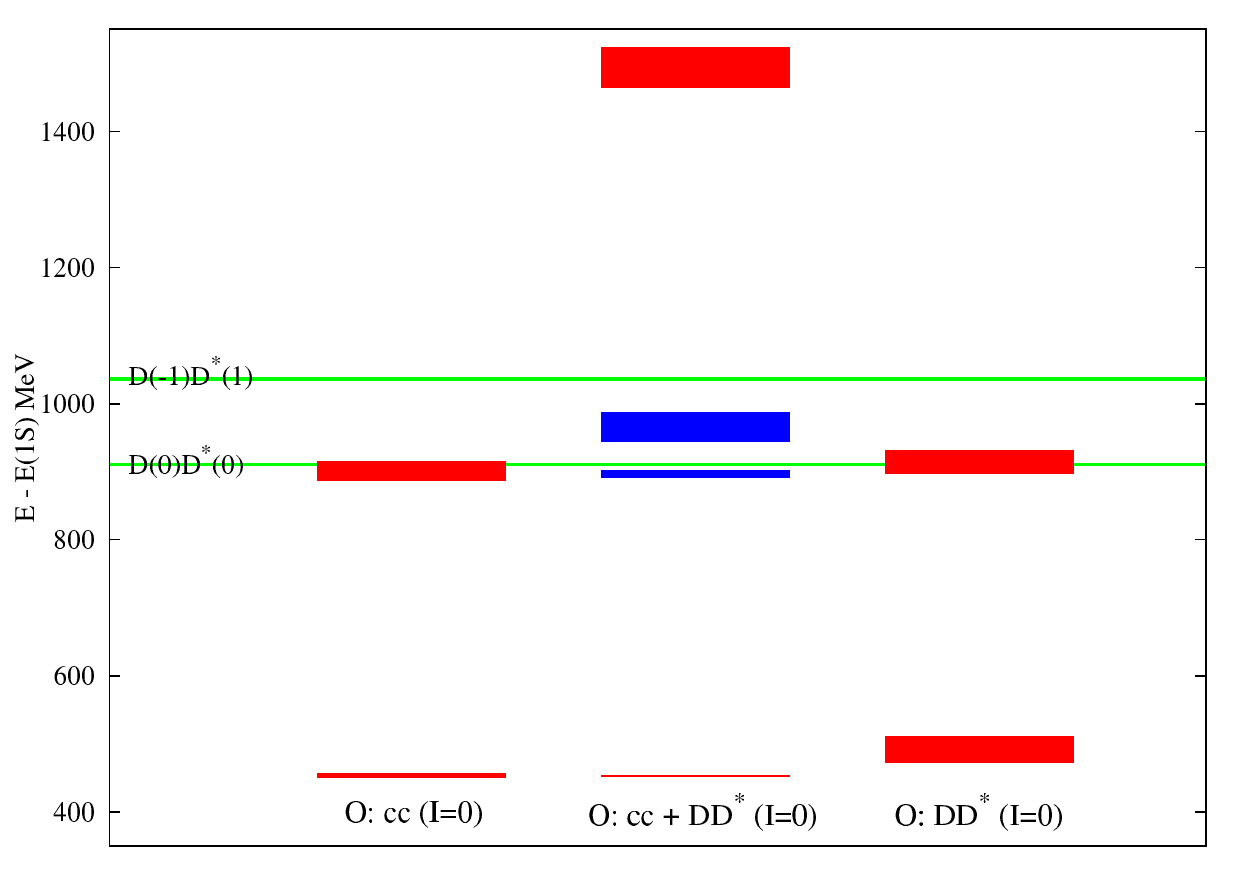}  
\caption{\label{fig:detar}   The spectra in the $Z_c^+$ channel with $I(J^{PC})=1(1^{+-})$ (upper pane), and in $X(3872)$ channel with $0(1^{++})$ (lower pane)  \cite{DeTar:lat14}. The red and blue boxes  are energies obtained  from the simulation for various interpolator basis indicated at the bottom. The $cc$ and $cc~\pi$  denote basis including charmonium and charmonium-pion interpolators, respectively.   
Lower pane: The lowest state is conventional $\chi_{c1}$, the  lower blue state is a candidate for $X(3872)$ with $I=0$, and  the green lines denote non-interacting energies of $D\bar D^*$. Upper pane: 
the green levels denote non-interacting $\psi\pi$ states, while the light blue levels denote $D\bar D^*$.  The state $D(1)\bar D^*(-1)$ is  not seen in both channels \cite{DeTar:lat14}. }
\end{center}
\end{figure} 
 
 \begin{figure}[htb] 
 \begin{center}
\includegraphics*[width=0.45\textwidth,clip]{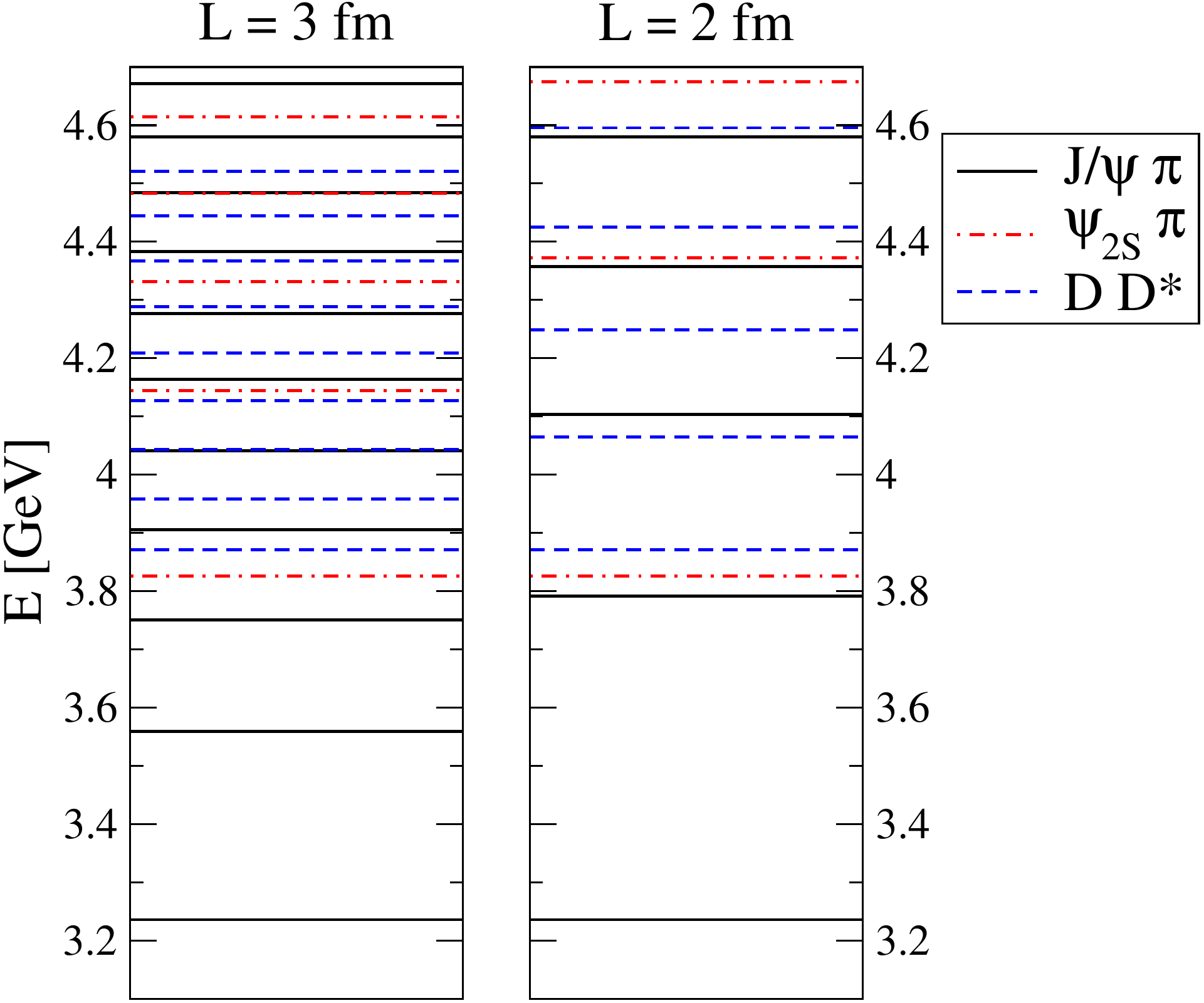}   $\quad $
 \includegraphics*[width=0.45\textwidth,clip]{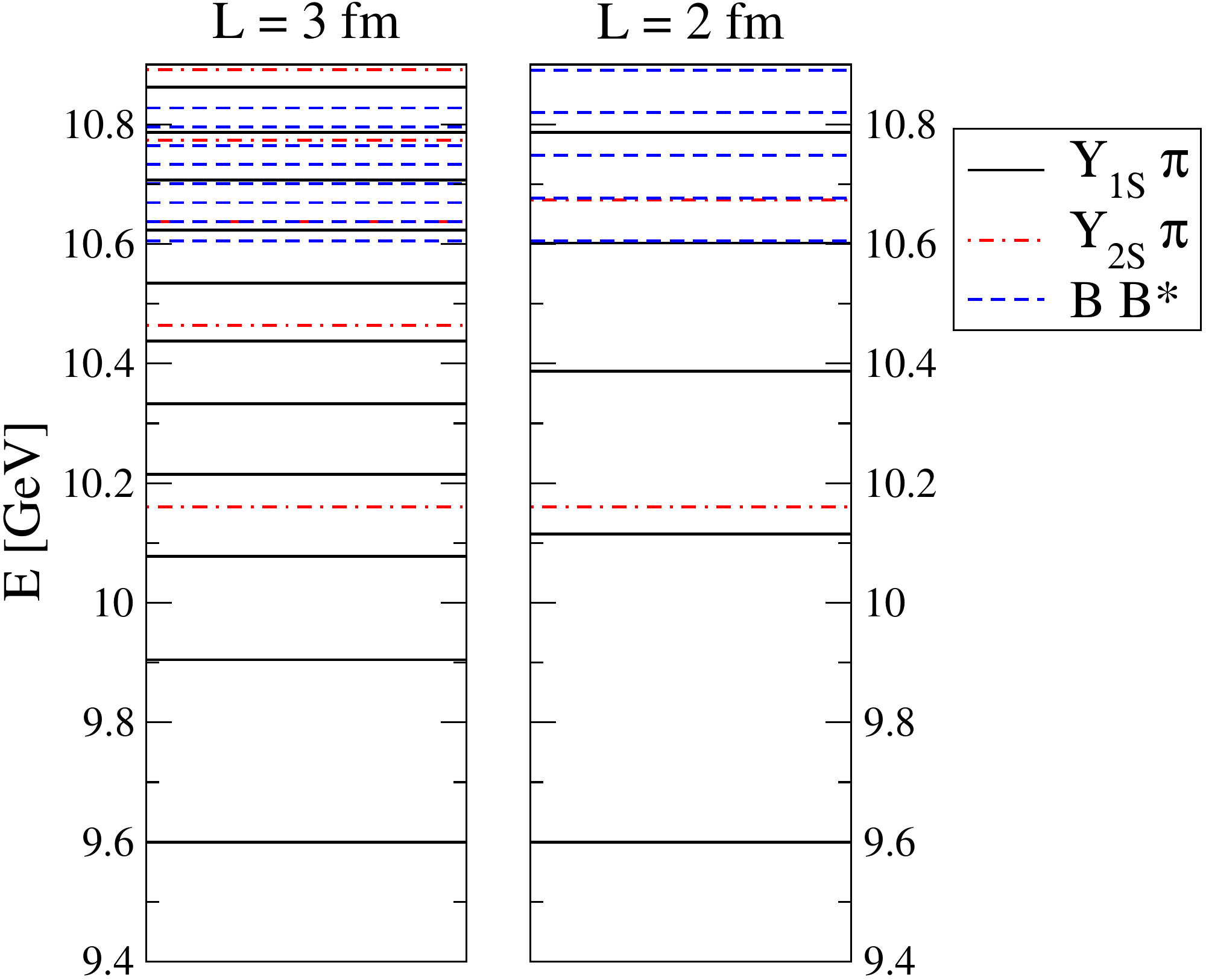}   
\caption{\label{fig:challenge}   The energies of the non-interacting two-meson states for physical $m_\pi$ and $L=2,~3~$ fm in the $Z_c^+$ (left) and $Z_b^+$ (right) channels with $J^{PC}=1^{+-}$.  Only three channels are shown and the spectrum is even denser when all channels are considered (see Fig. 2). Exotic experimental candidates are $Z_b^+(10610)$ and $Z_b^+(10650)$, while experimental $Z_c^+$ candidates are shown in  Fig. 2.   }
\end{center}
\end{figure} 

  \underline{$\mathbf{Z_c^+}$}:  Several charged-charmonia with quark content $\bar cc\bar d u$ were discovered recently in experiment. Most notably these are $Z^+(4430)$ with $I^G(J^{PC})=1^+(1^{+-})$ discovered by Belle and confirmed by LHCb, and $Z_c^+(3900)$ with unknown $J^P$ discovered by BESIII and confirmed by Belle and CLEOc  \cite{Brambilla:2014jmp}.   It is important to note that $Z_c^+(3900)$ was found in $J/\psi\,\pi$ invariant mass only through $e^+e^- \to Y(4260)\to  (J/\psi\, \pi^+)\pi^- $.  No resonant structure in $J/\psi\, \pi^+$ was seen in  $\bar B^0\to (J/\psi\, \pi^+) K^-$ by BELLE \cite{Chilikin:2014bkk}, in $\bar B^0 \to (J/\psi \pi^+) \pi^-$  by LHCb  \cite{Aaij:2014siy} or in  $\gamma p\to (J/\psi \,\pi^+) n$ by COMPASS    \cite{Adolph:2014hba}.     This might indicate that  the peak seen in $Y(4260)$ decay might not be of dynamical origin \cite{Chen:2013coa,Swanson:2014tra}.  
  
  The $Z^+(4430)$ lies close to $D^*\bar D_1$ threshold and the only lattice search for it  was performed in the quenched approximation soon after its discovery in 2007 \cite{Meng:2009qt}.    The attraction between $D^*$ and $\bar D_1$ was found using L\"uscher's approach, but no indication for  a bound state. Further lattice studies were  discouraged since this state lies above multiple thresholds. 
  
  All lattice searches for $Z_c^+$ considered $I^G(J^{PC})=1^+(1^{+-})$ channel, which is the most relevant experimentally. 
  The first lattice search for $Z_c^+(3900)$ considered $J/\psi\, \pi$ and $D\,\bar D^*$ scattering and only two-meson states   $J/\psi\, \pi$ and $D\,\bar D^*$ were found, but no additional candidate for $Z_c^+(3900)$ \cite{Prelovsek:2013xba}. 
  
 The  $D\bar D^*$ interpolators were used to determine the s-wave and p-wave phase shift  near $D\bar D^*$ threshold where experimental $Z_c(3900)$ is located.   Partially twisted boundary conditions were applied  taking into account s/p-wave  mixing  \cite{Chen:2014afa}. The authors conclude that  no evidence for $Z_c^+(3900)$ is found.   I would like to caution  that $D\bar D^*$ correlators relax to the true ground state  $e^{-(m_{J/\psi}+m_\pi)t}$ at large $t$ in \cite{Prelovsek:2013xba}, so the   energies   may not be reliable if only $D\bar D^*$ interpolators are employed. 
  
   The most extensive lattice search for  $Z_c^+$ with mass below $4.2~$GeV  was performed in \cite{Prelovsek:2014v2,Leskovec:lat14}.  The major challenge is presented by the two-meson states $J/\psi\, \pi$, $\psi_{2S}\,\pi$, $\psi_{1D}\,\pi$, $D\bar D^*$, $D^*\bar D^*$ and $\eta_c\,\rho$ that are  inevitably present in the  $I^G(J^{PC})=1^+(1^{+-})$ channel, in addition to  possible  $Z_c^+$ candidates.  
    Thirteen two-meson states are expected on the lattice with $L\simeq 2~$fm and $m_\pi\simeq 266~$MeV in the energy region of interest, and their non-interacting energies are shown by the horizontal lines in Fig. \ref{fig:Zc_prelovsek}.     
   The lattice spectrum of eigenstates is extracted using a number of  meson-meson  and  $[\bar c\bar d]_{3_c}[cu]_{\bar 3_c}$ interpolating fields. Black circles show the lowest thirteen energy levels that are identified as  two-meson states: they lie close to the non-interacting energies and show large overlap to the corresponding two-meson interpolators.  No additional state that could represent  $Z_c^+$  is found below $4.2~$GeV  \cite{Prelovsek:2014v2}. 
  The green state  at $4.4~$GeV  is not identified  with $Z_c^+$ since further two-meson states lie in this energy region.  
    
   \begin{figure}[htb] 
 \begin{center}
\includegraphics*[width=0.50\textwidth,clip]{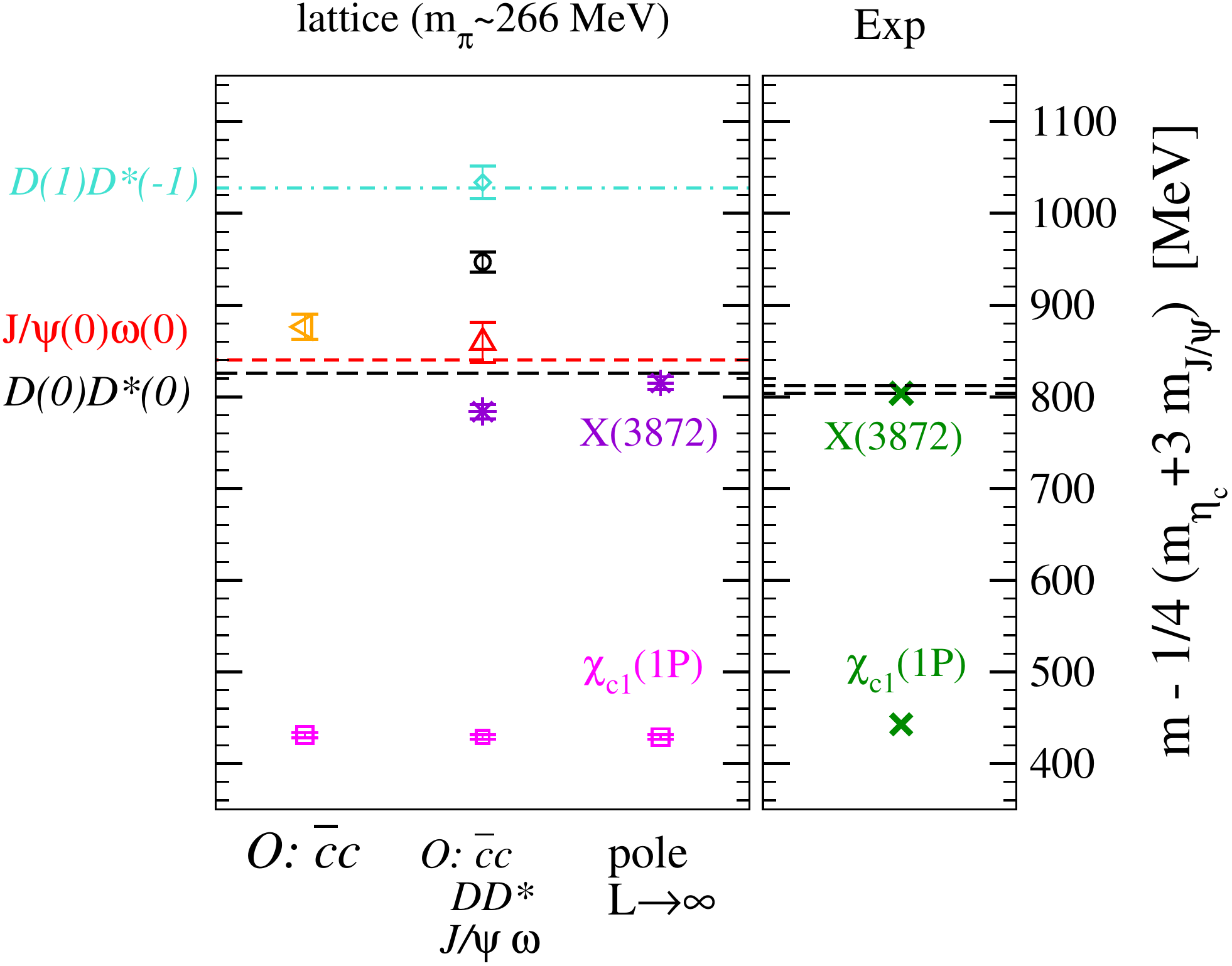}  
\caption{\label{fig:X_prelovsek}   The  spectrum in the $J^{PC}=1^{++}$, $I=0$ channel where charmonium-like $X(3872)$ resides: lattice results  for two choices of the interpolator basis (points) and the energies of the non-interacting two-meson states (lines) \cite{Prelovsek:2013cra}. The  phase shift analysis renders the locations of the poles in $D\bar D^*$ scattering, which  represent $\chi_{c1}$ and $X(3872)$ bound states. These are close to the experimental masses, shown on the right.  }
\end{center}
\end{figure}

   A preliminary study that does not incorporate all low-lying two-mesons states around $4.2~$GeV seems to render a $Z_c^+$ candidate at $4.16~$GeV \cite{Prelovsek:2014v1}\footnote{These preliminary results were reported at this meeting.}, which  is however not robust after further two-meson states around $4.2~$GeV are implemented \cite{Prelovsek:2014v2}.  Cautionary remarks  based on this study are provided in Section V of \cite{Prelovsek:2014v2}.
   
   The lattice simulation of the same channel based on the HISQ action gives the  energies in Fig. \ref{fig:detar} for various interpolator basis \cite{DeTar:lat14}. The figure illustrates that the two-meson states are seen if the corresponding interpolators are employed. The argument why $D(1)\bar D^*(-1)$ level does not appear is given in \cite{DeTar:lat14}. No additional state that could represent $Z_c^+$ is found.   
    
 A large number of two-meson states apparently represents a major challenge in searching for $Z_c^+$ or $Z_b^+$ candidates using lattice QCD. Their spectra becomes denser with growing $L$, which poses a serious challenge already at $L\simeq 3~$fm, as shown in Fig. \ref{fig:challenge}. 
    \\

\underline{$\mathbf{X(3872)}$}: The Belle discovery of neutral charmonium-like $X(3872)$ in 2013   initiated a new era for unconventional-hadron spectroscopy. The state has $J^{PC}=1^{++}$, it is located within $1~$ MeV of $D^0\bar D^{0*}$ threshold and decays to $I=0$ as well as $I=1$ final states.  The lattice simulations before 2013 used only $\bar cc$ interpolators and therefore ignored the effect of $D\bar D^*$ threshold. These typically rendered one energy level  near $D\bar D^*$ threshold, but one could not reliably establish whether this state corresponds to $X(3872)$ or $D(0)\bar D^*(0)$.

A candidate for the charmonium(like) state $X(3872)$ was found $11\pm 7~$MeV below the $D\bar D^*$ threshold for $J^{PC}\!=\!1^{++}$, $I\!=\!0$, $N_f\!=\!2$ and $m_\pi\!\simeq\! 266~$MeV  \cite{Prelovsek:2013cra} (see Fig. \ref{fig:X_prelovsek}). This was the first lattice simulation that establishes a candidate for $X(3872)$ in addition to $\chi_{c1}$  and the nearby scattering states $D\bar D^*$  and $J/\psi\,\omega$. The large and negative $a_0\!=\!-1.7\pm 0.4~$fm for $D\bar D^*$ scattering length   is one  indication for a shallow bound state $X(3872)$, as can be understood from  (\ref{ER}).  The mass of the $X$ is determined from the position of the pole in $S$-matrix which is obtained by interpolating $D\bar D^*$ scattering phase shift near threshold (\ref{B}, \ref{ER}). The established $X(3872)$ has a large overlap with $\bar cc$ as well as $D\bar D^*$ interpolating fields; it is absent if one or the other type of interpolators are omitted.  Only the $D\bar D^*$ and $J/\psi\,\rho$ scattering states are found in the $I\!=\!1$ channel, and there is no candidate for $X(3872)$ in this case \cite{Prelovsek:2013cra}. This is in agreement  with a popular interpretation that $X(3872)$ is dominantly $I\!=\!0$, while its small $I\!=\!1$ component arises solely from the isospin breaking and is therefore absent in the simulation with $m_u\!=\!m_d$.  

New  evidence for $X(3872)$ with $I\!=\!0$ came from the HISQ action at this meeting  \cite{DeTar:lat14} and it is represented by the lower blue state in Fig. \ref{fig:detar}. The candidate for $X(3872)$  is found in the spectrum only if both $\bar cc$ and $D\bar D^*$ interpolators are used. 

Recent analytic studies   consider the quark mass dependence, the volume dependence and the effect from the isospin breaking relevant for future lattice studies of $X(3872)$ \cite{Jansen:2013cba,Garzon:2013uwa,Albaladejo:2013zxa}.  \\

\underline{$\mathbf{X(4140)}$}: A structure called $X(4140)$ was found in the $J/\psi\;\phi$ invariant mass by  CDF and more recently by CMS  and D0, while its quantum numbers are  unknown experimentally \cite{Brambilla:2014jmp}. 

\begin{figure}[tb] 
 \begin{center}
\includegraphics*[width=0.4\textwidth,clip]{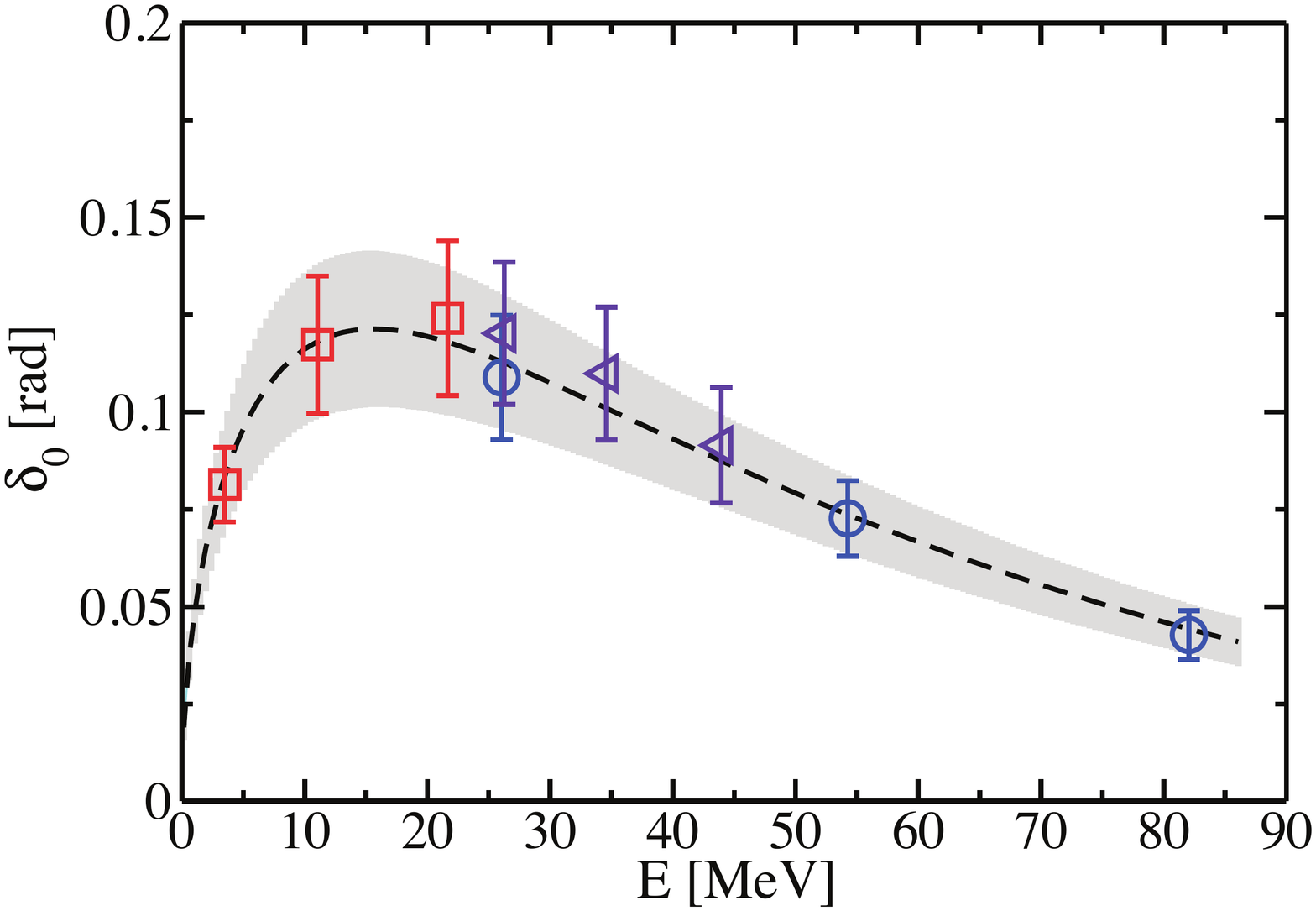}  $\quad$
\includegraphics*[width=0.40\textwidth,clip]{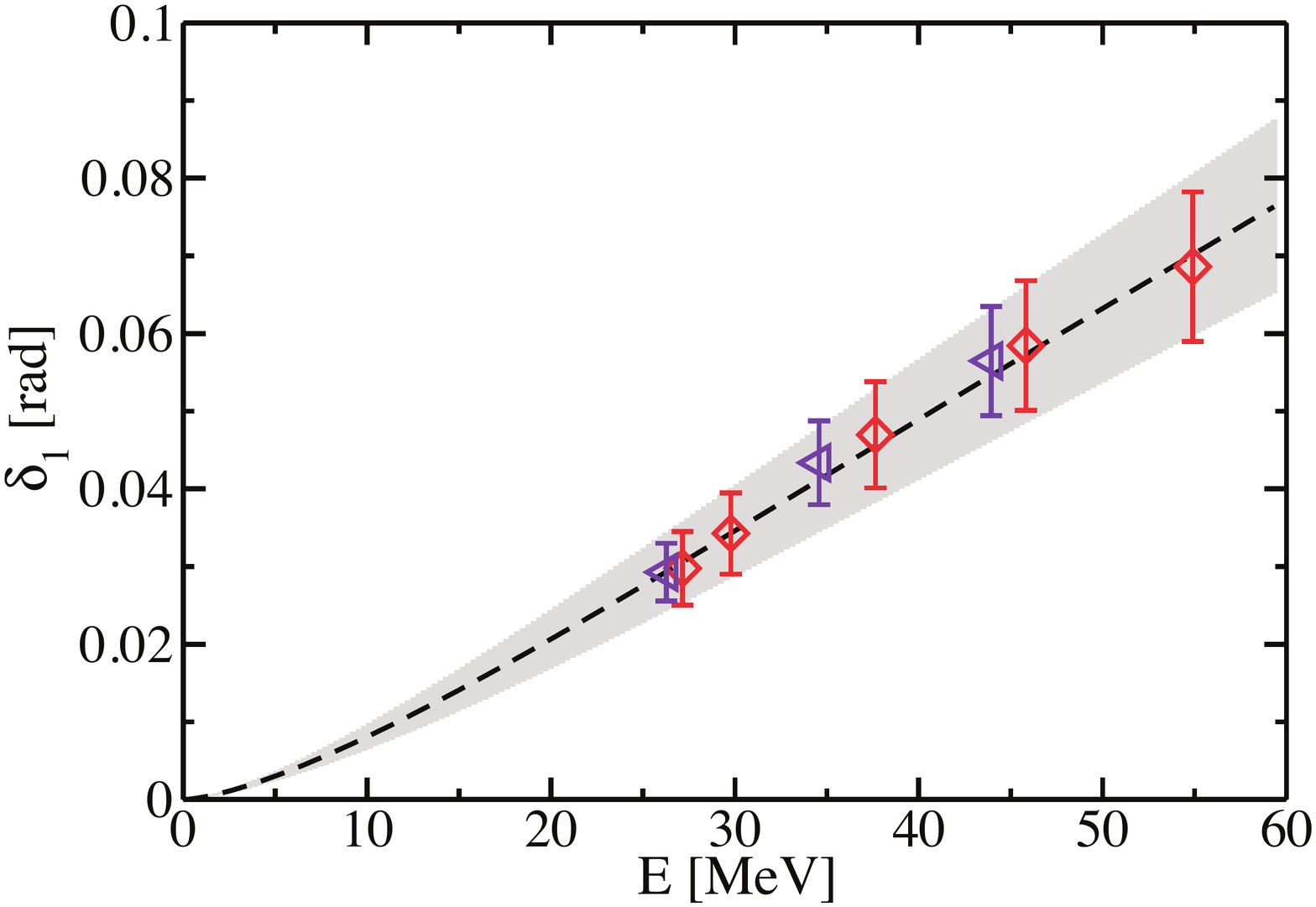}  
\caption{\label{fig:sasaki}  The s-wave and p-wave phase shift for $J/\psi\; \phi$ scattering  aimed at searching for $X(4140)$ \cite{Ozaki:2012ce}.     }
\end{center}
\end{figure}

The s-wave and p-wave $J/\psi\;\phi$ scattering phase shift in Fig. \ref{fig:sasaki} was extracted  in a $N_f=2+1$ simulation   using twisted boundary conditions \cite{Ozaki:2012ce}, where mixing was taken into account. The $\bar ss$ annihilation contribution was omitted.   The phase shifts do not support a resonant structure. \\

\begin{figure}[tb] 
 \begin{center}
\includegraphics*[width=0.45\textwidth,clip]{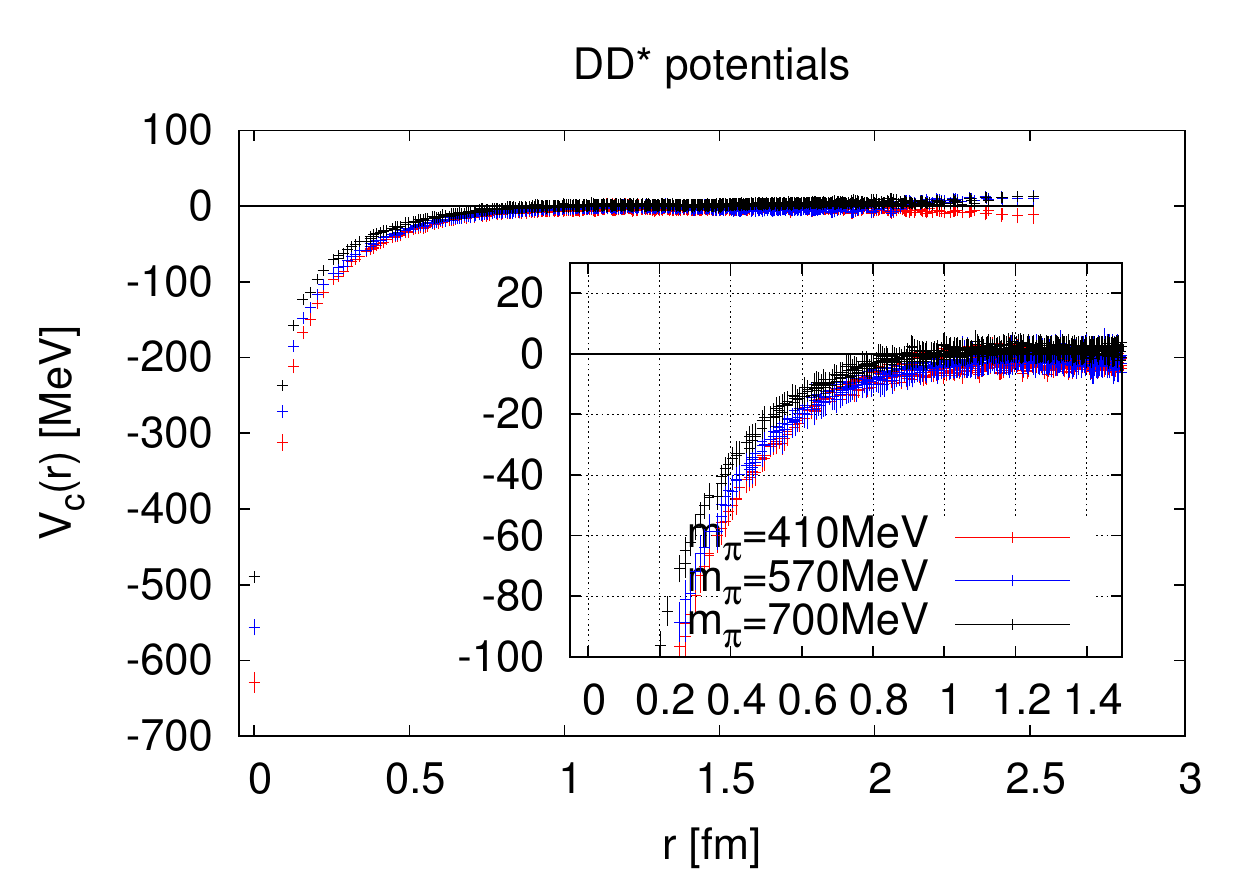}  $\quad$
\includegraphics*[width=0.45\textwidth,clip]{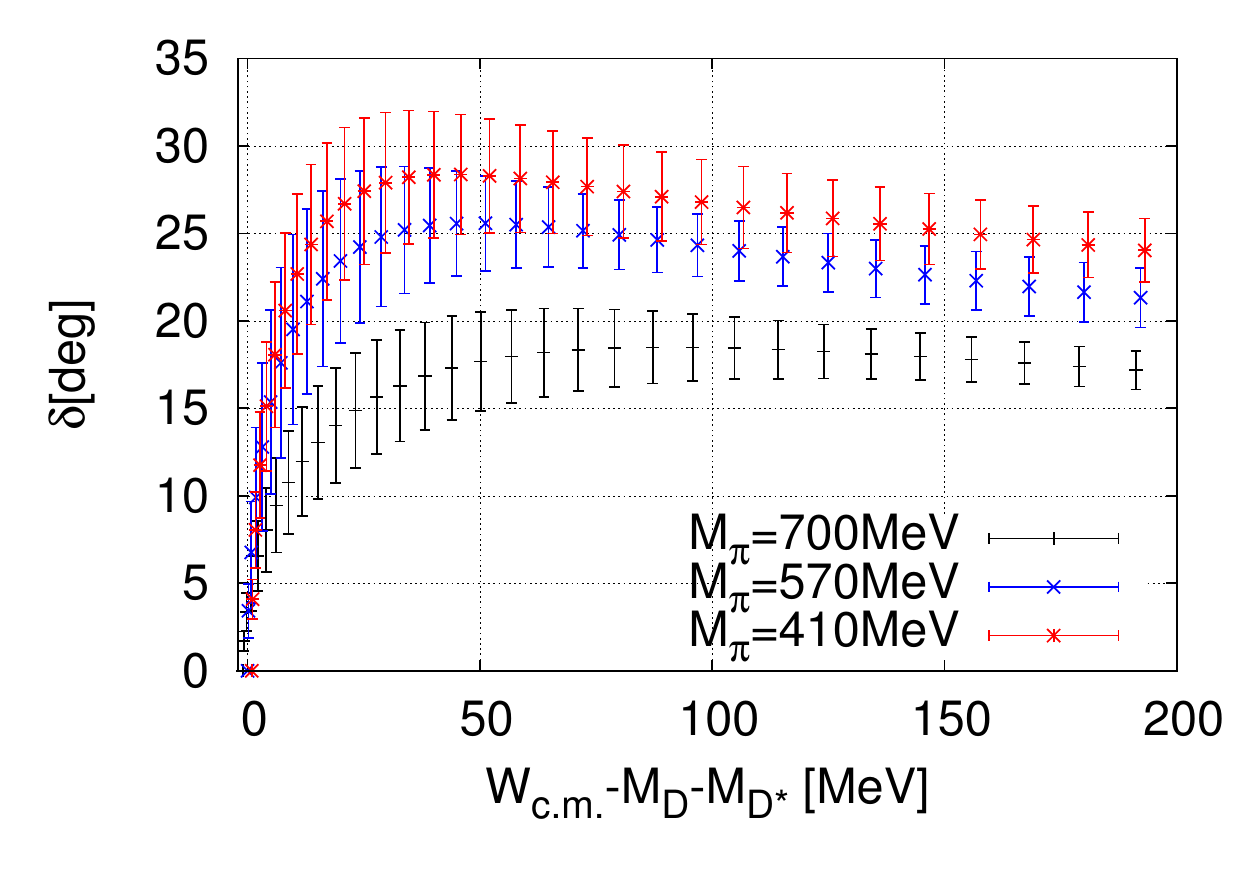}  
\caption{\label{fig:halqcd}  The potential $V(r)$ between $D$ and $D^*$ mesons as a function of distance $r$ calculated using HALQCD method, and the resulting $DD^*$ phase shift for this double-charm channel \cite{Ikeda:2013vwa}.     }
\end{center}
\end{figure}

\underline{$\mathbf{cc\bar d\bar u}$ {\bf tetraquarks}}:
The potential $V(r)$ between $D\!=\!\bar uc$ and $D^*\!=\!\bar dc$ as a function of  distance $r$ was extracted using the HALQCD time-dependent method \cite{HALQCD:2012aa}. The resulting potential in  Fig. \ref{fig:halqcd} is attractive, but the corresponding $DD^*$ scattering phase shift does not start at $\delta(0)=\pi$, which indicates that there is no $cc\bar d\bar u$ tetraquark bound state at the simulated $m_\pi$ \cite{Ikeda:2013vwa}. 

No exotic candidates were seen  also in a preliminary study of the double-charm tetraquark channel with $I=0$ and $I=1$  \cite{Guerrieri:lat14} that has been presented at this meeting.  \\

\begin{figure}[tb] 
\begin{center}
\includegraphics*[width=0.80\textwidth,clip]{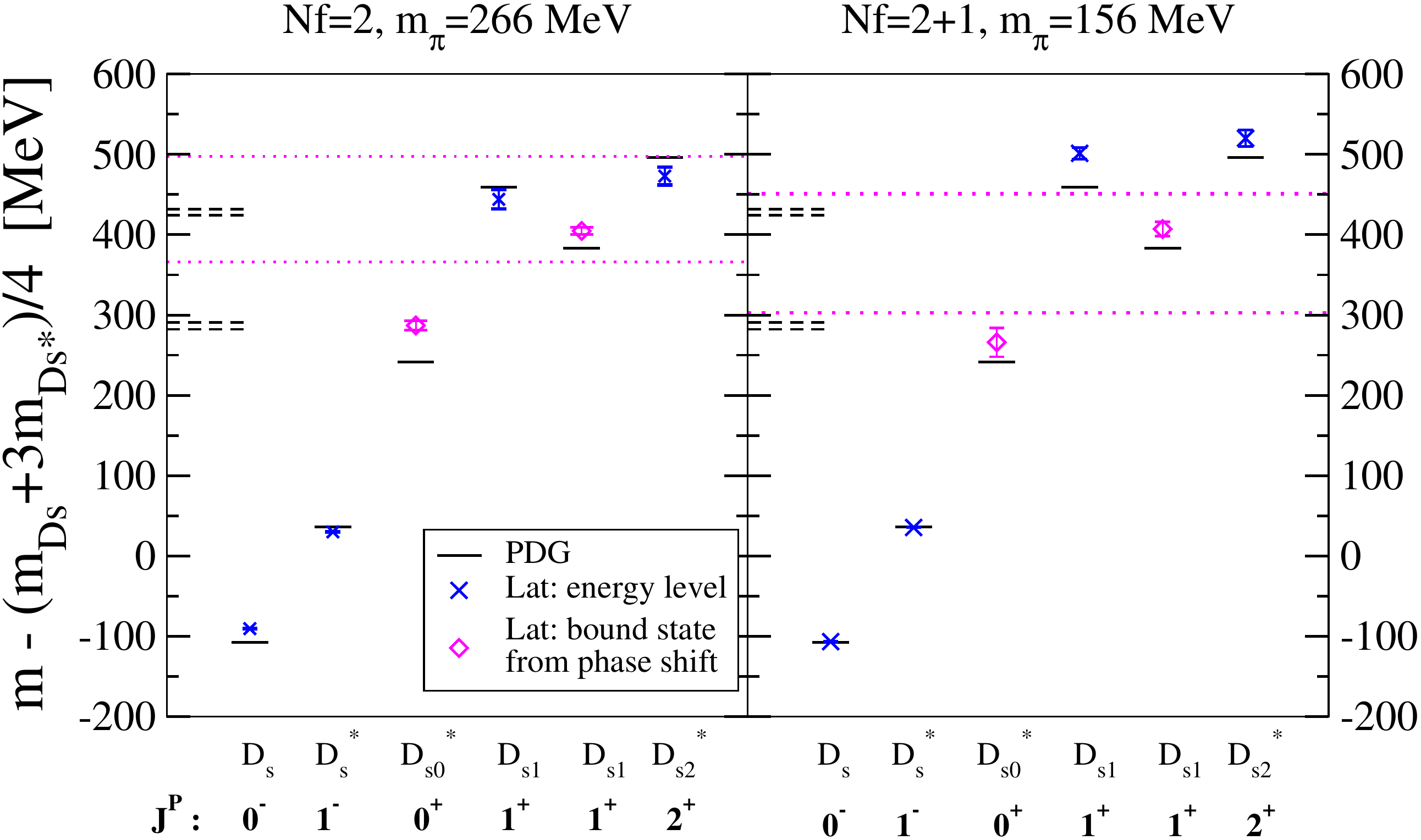}  
\caption{\label{fig:Ds}   Spectra of $D_s$ mesons on two ensembles: the magenta points indicate masses inferred from poles in $D^{(*)}K$ scattering matrix; blue points are masses obtained using the single-hadron approach  \cite{Mohler:2013rwa,Lang:2014yfa,Lang:lat14}. The back lines represent experimental masses.  }
\end{center}
\end{figure} 

\underline{\bf $\mathbf{D_{s0}^*}~$, $\mathbf{D_{s1}}$}: The quark models expected $D_{s0}^*(2317)$ and $D_{s1}(2460)$  above  $DK$ and $D^*K$ thresholds, but they were experimentally found slightly below them.  These are one of few shallow  bound states in the meson sector, and therefore deserve special attention. Many previous simulations extracted the ground states with $J^P=0^+$ or $1^+$ using $\bar sc$ interpolators, but one could  not reliably establish whether these correspond to $D_{s0}/D_{s1}$ mesons or $D^{(*)}(0)K(0)$ states.  

The first lattice simulation that takes the effect of these thresholds into account was performed during past two years  \cite{Mohler:2013rwa,Lang:2014yfa,Lang:lat14}. It employs $DK$ and  $D^*K$ interpolating fields in addition to the $\bar sc$.  The position of thresholds is almost physical on $N_f\!=\!2+1$ ensemble with nearly physical  $m_\pi\simeq 156~$MeV. The $D^{(*)}K$ phase shift is extracted from each energy level and then parametrized in the region close to threshold using effective range formula (\ref{ER}).   The  scattering lengths are large and negative, which  further supports the existence of the shallow bound states.  Their binding momenta are obtained by searching for the pole in scattering matrix, which is realized at  $\cot(\delta(p_B))=i$ (\ref{B},\ref{ER}).   These poles are related to $D_{s0}^*(2317)$  and $D_{s1}(2460) $ bound states and are close to the experimental masses.     The summary of the resulting $D_s$ spectrum for these two  as well as other $D_s$ states is shown in Fig. \ref{fig:Ds}  for two values of pion masses. 

A preliminary  spectrum for these channels using $DK$,  $\bar sc$ as well as $D_s\eta$ interpolating fields  was presented at this conferences for  zero  and non-zero total momenta \cite{Ryan:lat14}. The goal is to determine the  scattering matrix for the coupled channel system $DK - D_s\eta$, and look for signatures of  bound states and resonances in it. \\

\underline{{\bf Baryon} $\Lambda(\mathbf{1405})$}: Experimentally, the $\Lambda(1405)$ baryon is a resonance in $\Sigma \pi$ scattering located slightly below $KN$ threshold.

 \begin{figure}[tb] 
 \begin{center}
 \includegraphics*[width=0.60\textwidth,clip]{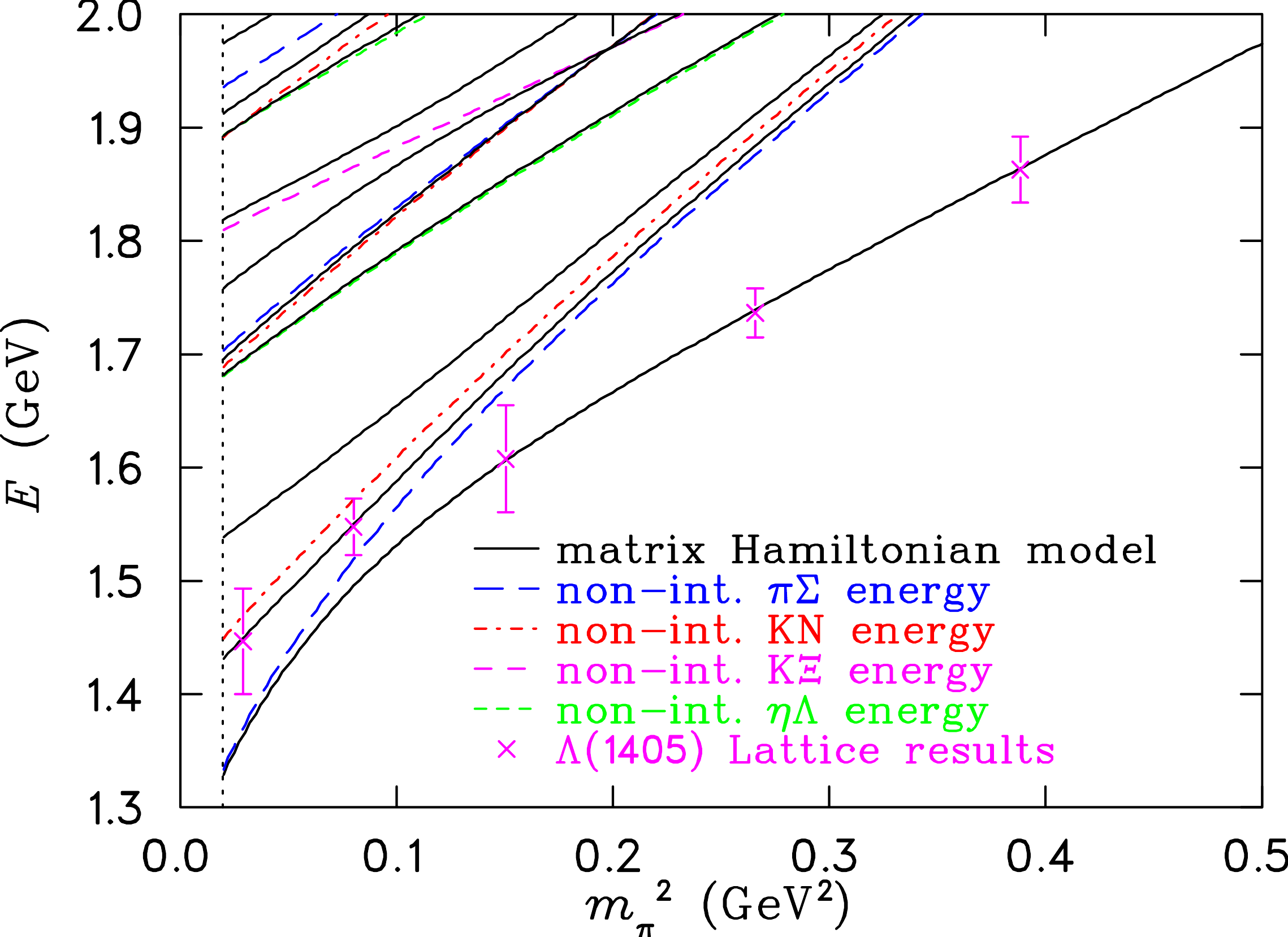}     
\caption{\label{fig:spectrum_adelaide}  The study of $\Lambda(1405)$ channel: the lines represent the eigenvalues of the finite volume EFT for the coupled system of $uds$, $\pi\Sigma$, $\bar KN$, $K\Xi$, which is fitted to  the ground state energy obtained using $uds$ interpolators (pink points)  \cite{Leinweber:lat14}.  }
\end{center}
\end{figure} 

 \begin{figure}[tb] 
 \begin{center} 
  \includegraphics*[width=0.60\textwidth,clip]{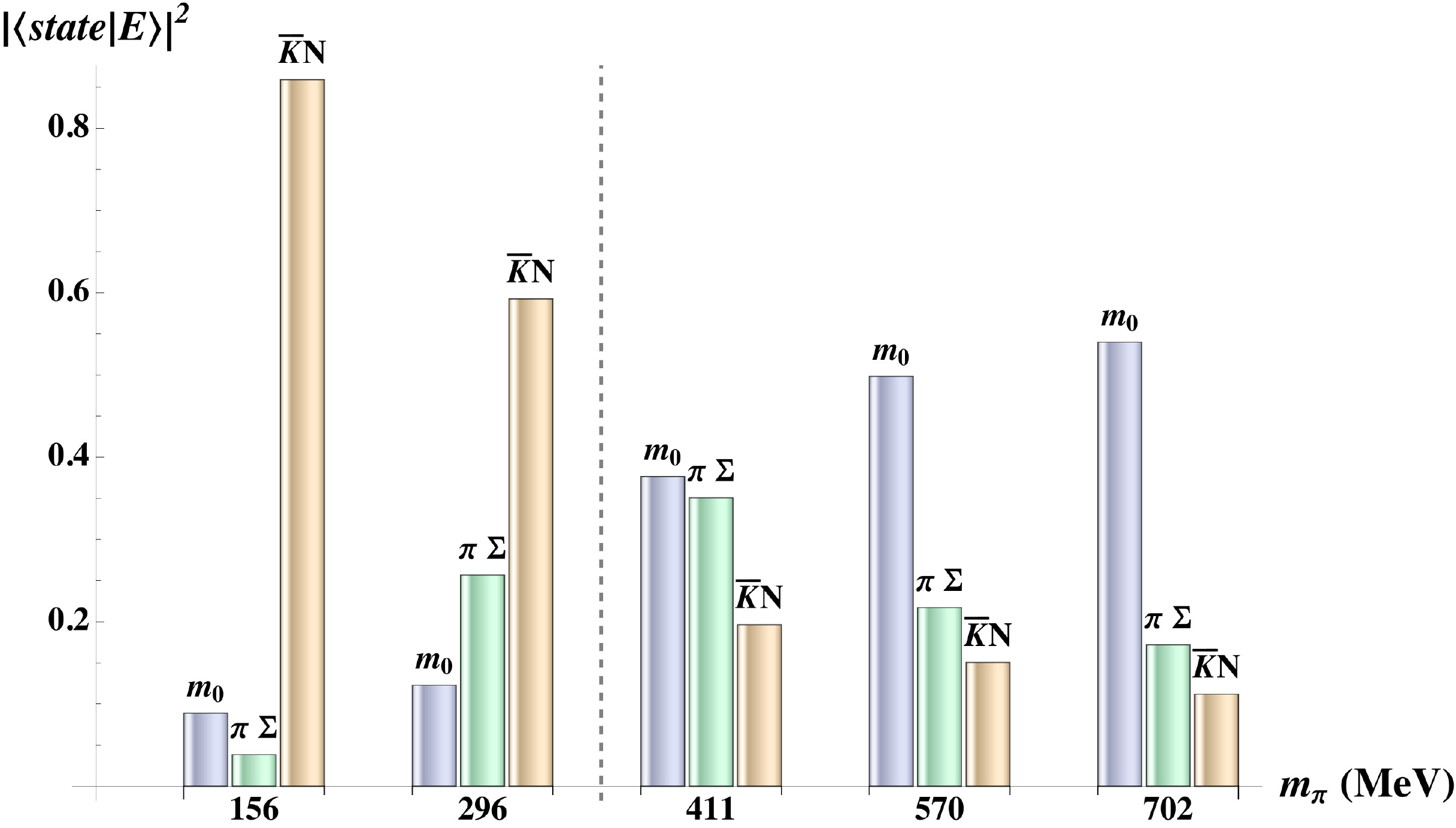}  
\caption{\label{fig:composition_adelaide}   Composition of the $\Lambda(1405)$  eigenstate inferred from EFT in terms of various Fock components  $uds$ (denoted by $m_0$), $\pi\Sigma$ and $\bar KN$  \cite{Leinweber:lat14}.}
\end{center}
\end{figure}

The ground state energies for a range of $m_\pi$ (pink levels in Fig. \ref{fig:spectrum_adelaide}) are obtained on PACS-CS ensembles using ${\cal O}=uds$  and  fitted with the eigenvalues of the Effective Hamiltonian Theory \cite{Leinweber:lat14}. This effective theory describes the interaction between $uds$, $\pi\Sigma$, $\bar KN$, $K\Xi$. Figure \ref{fig:spectrum_adelaide} shows the resulting eigenvalues of EFT (lines) and the composition of the $\Lambda(1405)$ eigenstate (histograms).  This state is found to be dominated by $\bar KN$ Fock component at the lightest $m_\pi$. The conclusions of this study are based on the simulation without $\Sigma\pi$ and $\bar KN$ interpolators, and need to be verified with a simulation including them. Note that the ground state energy can get modified with the inclusion of two-hadron interpolators, for example  in the case of a resonance \cite{Mohler:2012na,Lang:2012db} or near-threshold state \cite{Mohler:2013rwa}.

Preliminary results from including the  baryon-meson interpolating fields in the correlation matrix for the spectroscopy of $\tfrac{1}{2}^+$ nucleon channel  where presented at this meeting \cite{Kamleh:lat14,Verduci:lat14}.  The first results in the $\tfrac{1}{2}^-$ sector have been presented   in  \cite{Lang:2012db}.

\section{Rigorous treatment of hadronic resonances}

The rigorous treatment of a resonance in an elastic channel   $M_1M_2$ amounts to a determination of the discrete spectrum including two-meson states, determination of the scattering phase shift from each energy level and making a Breit-Wigner type fit  of the phase shift (\ref{R}), as described in Section 2.  The only hadron resonance studied in this way until recently is $\pi\pi\to\rho\to \pi\pi$, which has been simulated by a number of lattice collaborations until now (see for example \cite{Dudek:2012xn}). In the following I summarize results for other channels, where only first pioneering steps  have been made. \\
 
\begin{figure}[htb] 
 \begin{center}
(a) \includegraphics*[width=0.95\textwidth,clip]{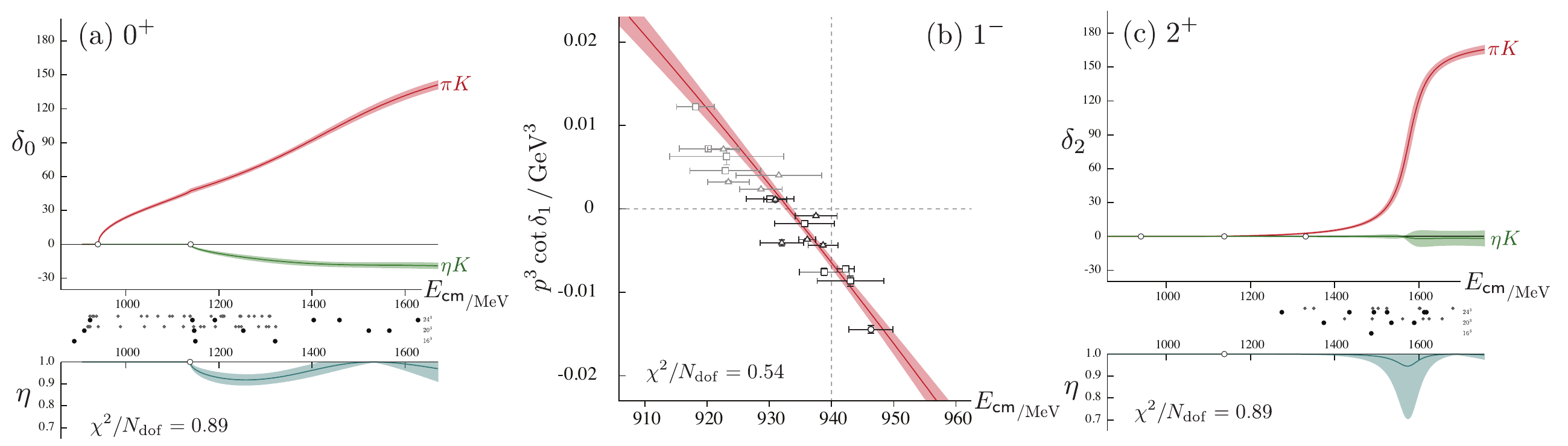}  $\quad$

\vspace{0.5cm}

(b) \includegraphics*[width=0.4\textwidth,clip]{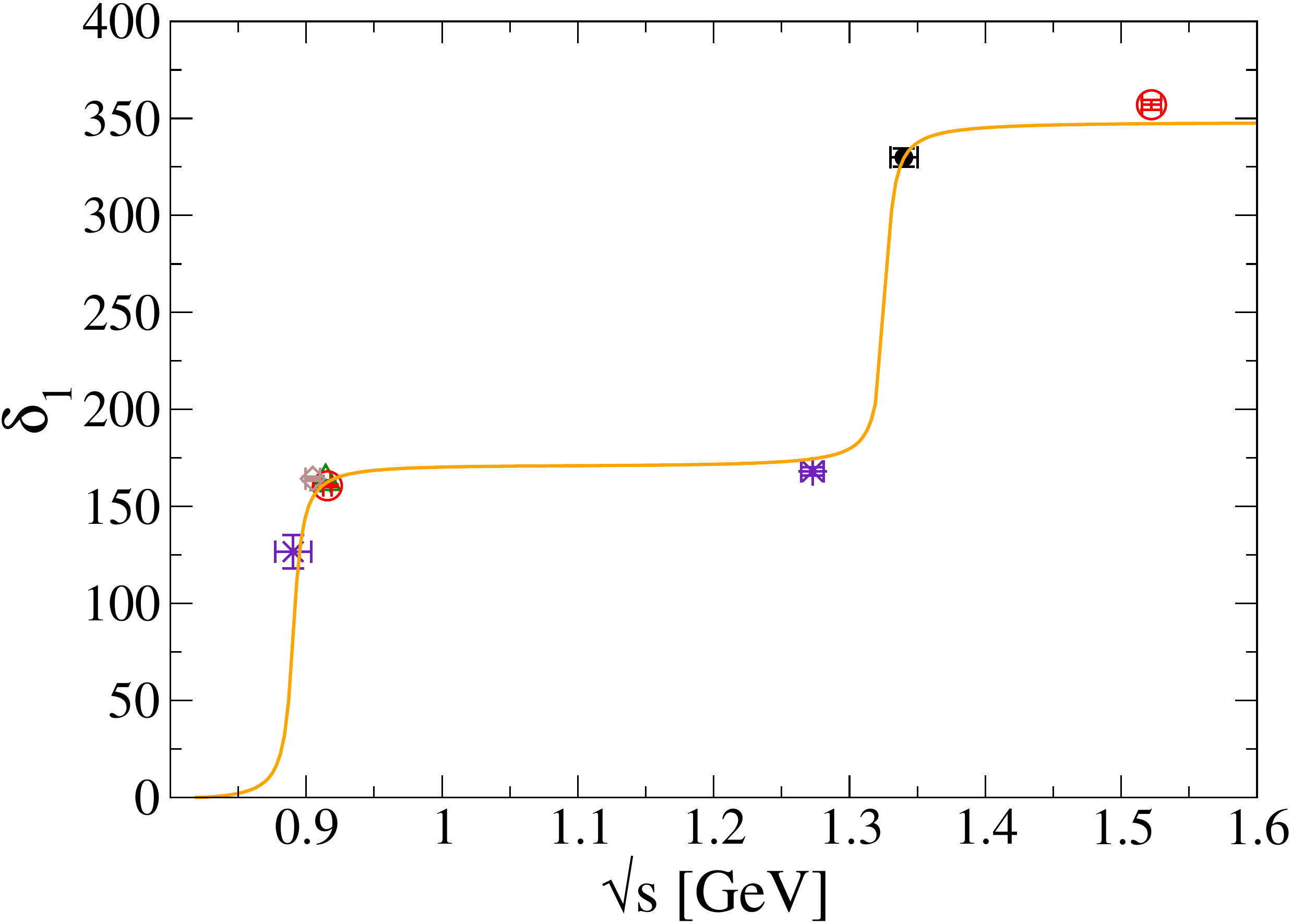}  $\quad$
(c) \includegraphics*[width=0.49\textwidth,clip]{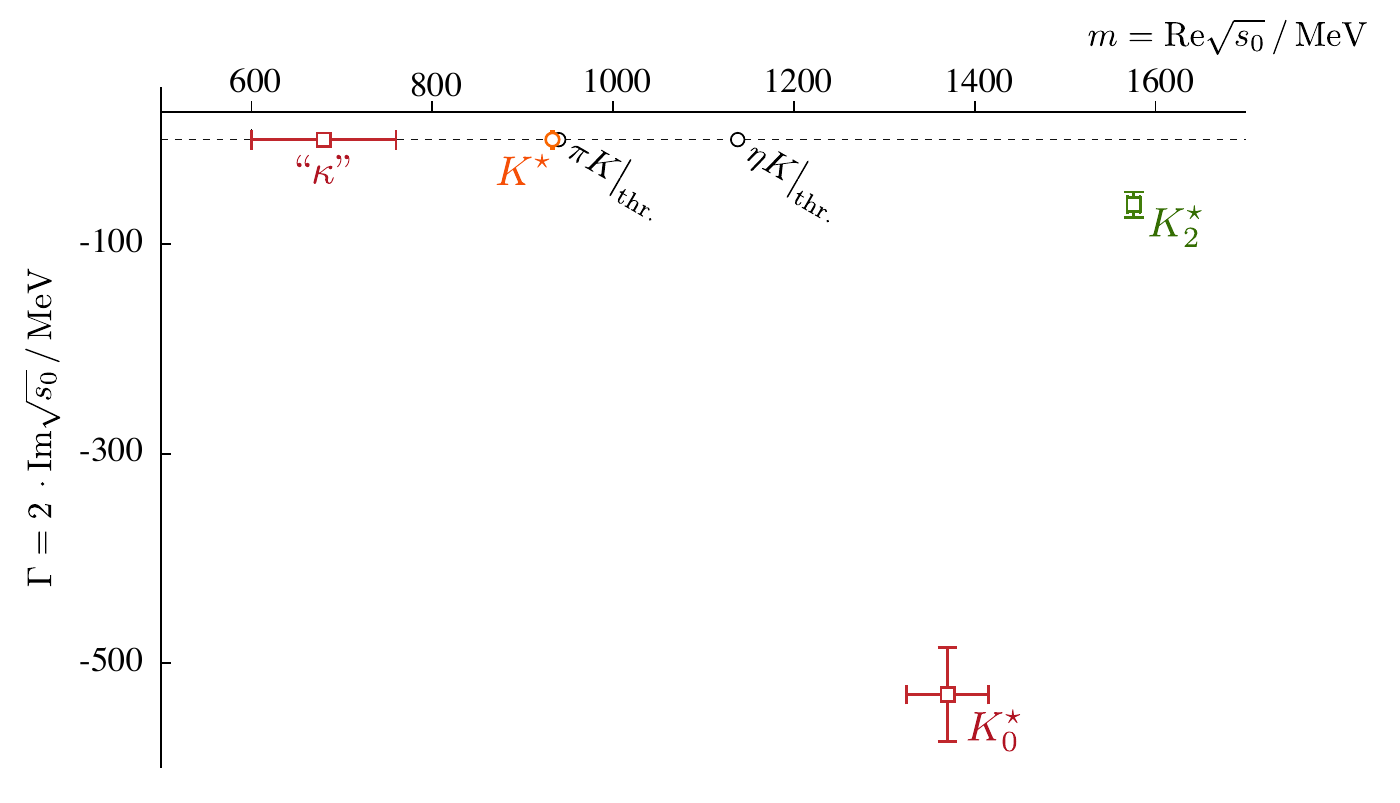}  
\caption{\label{fig:Kpi}   (b) $\delta_{l=1}^{K\pi}$ phase shift from  \cite{Prelovsek:2013ela}; (a) $\delta^{K\pi}$, $\delta^{K\eta}$ and the inelasticity $\eta$ from \cite{Dudek:2014qha,Wilson:lat14}; (c)    the poles of the scattering matrix in the complex plane  correspond to the strange mesons \cite{Dudek:2014qha,Wilson:lat14}. }
\end{center}
\end{figure} 

\begin{figure}[h] 
 \begin{center}
\includegraphics*[width=0.60\textwidth,clip]{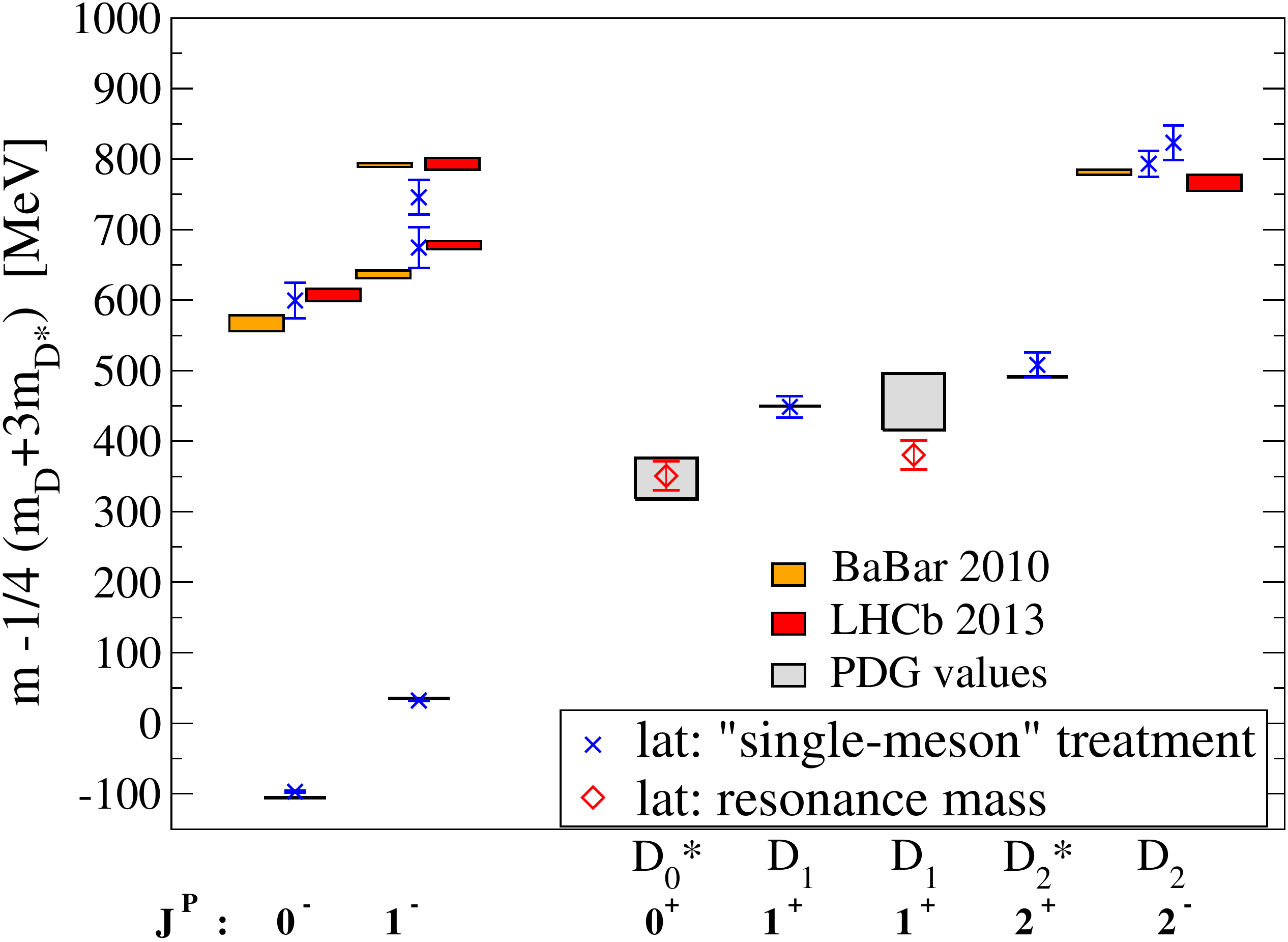}    
\caption{\label{fig:D_prelovsek}  The spectrum of $D$ mesons: lattice results for the masses in $0^+$ and $1^+$ channels are obtained using the Breit-Wigner fit of $D^{(*)}\pi$ phase shift (red diamonds); other masses rely on the single hadron approach (blue crosses)  \cite{Mohler:2012na}.  BaBar and more recent LHCb results are shown side-by-side.      }
\end{center}
\end{figure} 

\underline{$\mathbf{K^*}$, $\mathbf{\kappa}$, $\mathbf{K_0^*}$  {\bf and } $\mathbf{K_2}$}: $K^*$ mesons and in particular the $K^*(892)$ were frequently addressed in lattice simulations, but always ignoring that the $K^*(892)$  decays strongly. 
 
The $K\pi$ system with total momenta $P=\tfrac{2\pi}{L}e_z~,\ \tfrac{2\pi}{L}(e_x+e_y)$ and $0$ was simulated in \cite{Prelovsek:2013ela}, allowing the extraction of  the phase shift for p-wave scattering with $I\!=\!1/2$.  A Breit-Wigner fit of the phase in Fig. \ref{fig:Kpi}b renders a $K^*(892)$ resonance mass $m^{lat}=891\pm 14~$MeV and the $K^*(892)\to K\pi$ coupling $\tilde g^{lat}=5.7\pm 1.6$ compared to the experimental values $m^{exp}\approx 892$ MeV and  $\tilde g^{exp}=5.72\pm 0.06$, where $\tilde g$ parametrizes the $K^*\to K\pi$ width (\ref{R})\footnote{The couplings $\tilde g$ and $g$ (\ref{R}) are related by $g^2=\tilde g^2/(6\pi)$.}. Mixing of $p$ and $d$-wave is taken into account 
when extracting the phase shift around the $K^*(1410)$ and $K_2^*(1430)$ resonances. 
The $K^*(1410)$ resonance mass is found at $m^{lat}=1.33\pm 0.02 ~$GeV compared to $m^{exp}=1.414\pm 0.0015~$GeV assuming the experimental $K^*(1410)\to K\pi$ coupling.    

 The first simulation of two-coupled channel system $K\pi-K\eta$ was presented just before this meeting \cite{Dudek:2014qha,Wilson:lat14}.  
The scattering matrix in the complex plane was parametrized as a function of $E^{cm}$ and the parameters were extracted using the fit to the finite volume spectrum via L\"uscher-type method.  The phase shifts $\delta^{K\pi}$, $\delta^{K\eta}$  and the inelasticity $\eta$ corresponding to the scattering matrix $T_{ii}\propto \eta e^{2i\delta_{i}(p)}-1$  are shown in Fig. \ref{fig:Kpi}a.  The poles related to $\kappa$ and $K^*(892)$ are found below $K\pi$ threshold for the employed $m_\pi\simeq 400~$MeV, while $K_0^*(1430)$ and $K_2$ are found as resonances above threshold (see Fig. \ref{fig:Kpi}c). 
\\

\underline{$\mathbf{D_{0}^*}$ {\bf and} $\mathbf{D_{1}}$}:  The first rigorous simulation of a hadronic resonance that contains charm quarks addresses the  broad scalar $D_0^*(2400)$ and the axial $D_1(2430)$ charmed-light mesons    \cite{Mohler:2012na}. These appear in  s-wave scattering for  $D\pi$ and $D^*\pi$, respectively.  The resonance parameters are obtained using a Breit-Wigner fit to the elastic phase shifts at $N_f=2$ and $m_\pi\simeq 266~$MeV. The resulting $D_0^*(2400)$  mass is $351\pm 21~$MeV above the spin-average $\tfrac{1}{4}(m_D+3m_{D^*})$, in agreement with the experimental value of $347\pm 29~$MeV. The resulting $D_0^*\to D\pi$ coupling $g^{lat}=2.55\pm 0.21~$GeV is close to the experimental value $g^{exp}\le1.92\pm 0.14~$GeV, where $g$ parametrizes the width (\ref{R}). The results for $D_1(2430)$ are also found close to the experimental values; these are obtained by appealing to the heavy quark limit, where the neighboring resonance $D_1(2420)$ is narrow. These resonance masses  are compared to the experimental ones in Fig. \ref{fig:D_prelovsek}. The plot displays also the spectra for other charmed mesons  obtained within single-hadron approach. 
 
 It is puzzling why the strange $D_{s0}^*(2317)$ and the non-strange $D_0^*(2400)$ charmed scalar mesons have a mass within 1 MeV of each other experimentally, while one would naively expect a larger slitting $O(m_s)$. The question is whether this near degeneracy is due to the strange meson being pushed down or the non-strange one being pushed up. This puzzle can be addressed by considering the lattice results for $D_0^*(2400)$  \cite{Mohler:2012na}, which is found as a resonance in $D\pi$, and $D_{s0}^*(2317)$ \cite{Mohler:2013rwa}, which is found as a pole below $DK$.  Both masses are found close to experiment. This favors the interpretation that the near degeneracy is a consequence of strange meson being pushed down due to $DK$ threshold. On the other hand, the interpretation that $D_0^*(2400)$ is pushed up due to tetraquark Fock component $\bar u\bar ss c$  is disfavored since $N_f=2$ simulation \cite{Mohler:2012na} renders its mass close to the experiment without any strange content in the valence or sea sectors.

A preliminary spectrum for $I=1/2$ channel  channel   using $D\pi$,  $\bar uc$ as well as $D\eta$ interpolating fields  was presented at this   meeting  \cite{Ryan:lat14}. The  coupled-channel analysis  to extract the scattering parameters and resonances  is planned. 
 \\

\underline{$\mathbf{a_1}$ {\bf and} $\mathbf{b_1}$}: The light axial-vector resonances $a_1(1260)$ and $b_1(1235)$ were  explored for $N_f\!=\!2$  by simulating the corresponding scattering channels $\rho \pi$ and $\omega\pi$ \cite{Lang:2014tia}. 
Interpolating fields $\bar qq$ and  $\rho\pi$ or $\omega\pi$ were used to extract the $s$-wave phase shifts for the first time. It is  assumed that $\rho$ and $\omega$ are stable, which is  justified  in the energy region of interest for the employed  parameters $m_\pi\simeq 266~$MeV and $L\simeq 2~$fm  \cite{Roca:2012rx}.
A Breit-Wigner fit  of the $s$-phase shift gives the $a_1(1260)$ resonance mass  $m_{a_1}^{\textrm{res}}=1.435(53)(^{+0}_{-109})~$GeV compared to $m_{a_1}^{\textrm{exp}}=1.230(40)~$GeV. The $a_1$ width is parametrized in terms of the coupling $g$ (\ref{R}), which results in  $g=1.71(39)~$GeV compared to  $g^{\textrm{exp}}=1.35(30)~$GeV derived from experiment. 

The analytic framework for the scattering of unstable particles  in the finite volume was recently formulated in  \cite{Roca:2012rx}. The approach was applied to study the effect of the $\rho \to \pi\pi$ decay to the  $\rho\pi$ scattering, and this is found to be negligible for $L$ and $m_\pi$ employed in \cite{Lang:2014tia}.  \\

\begin{figure}[tb] 
 \begin{center}
\includegraphics*[width=0.49\textwidth,clip]{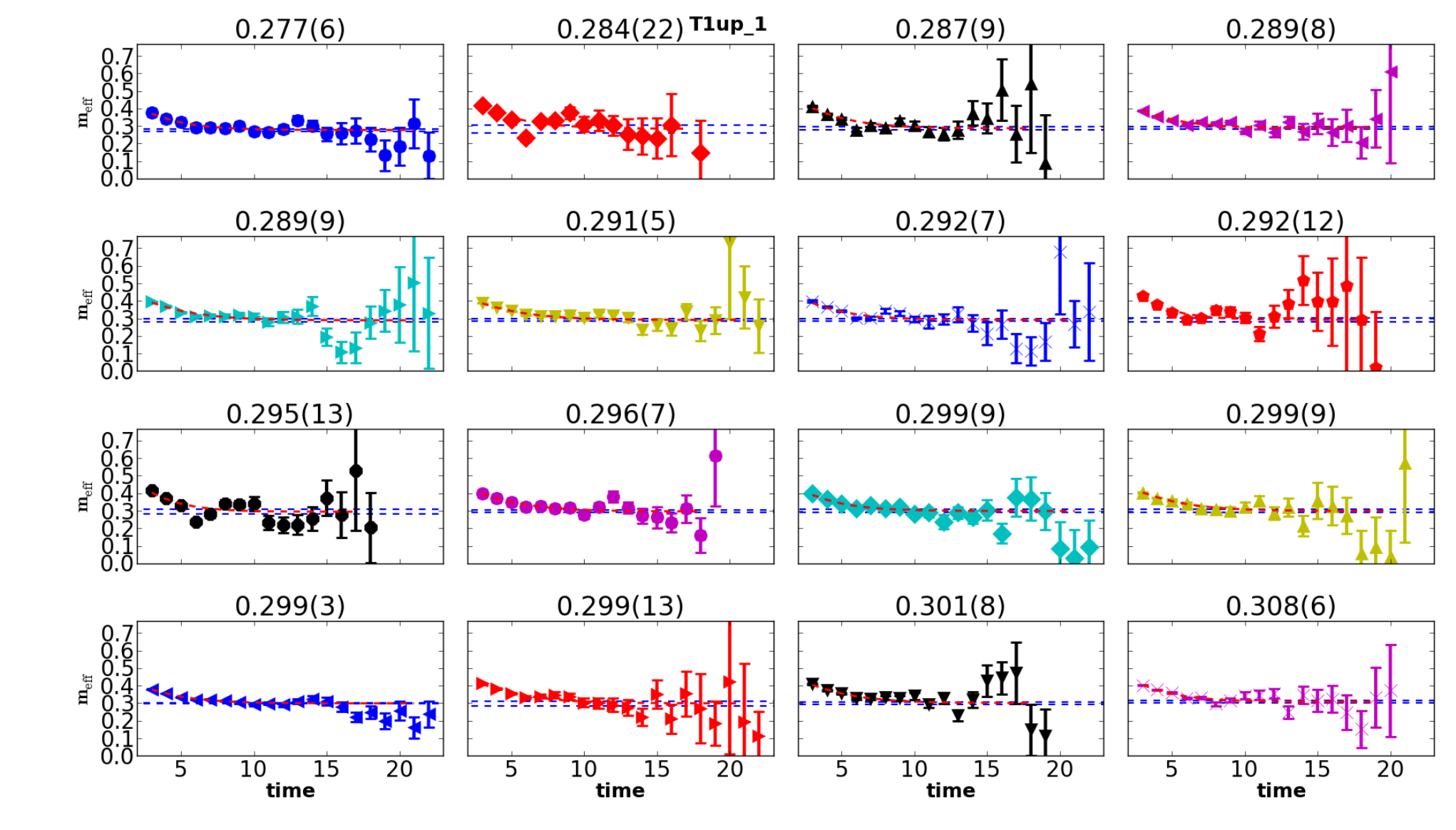}   
\includegraphics*[width=0.49\textwidth,clip]{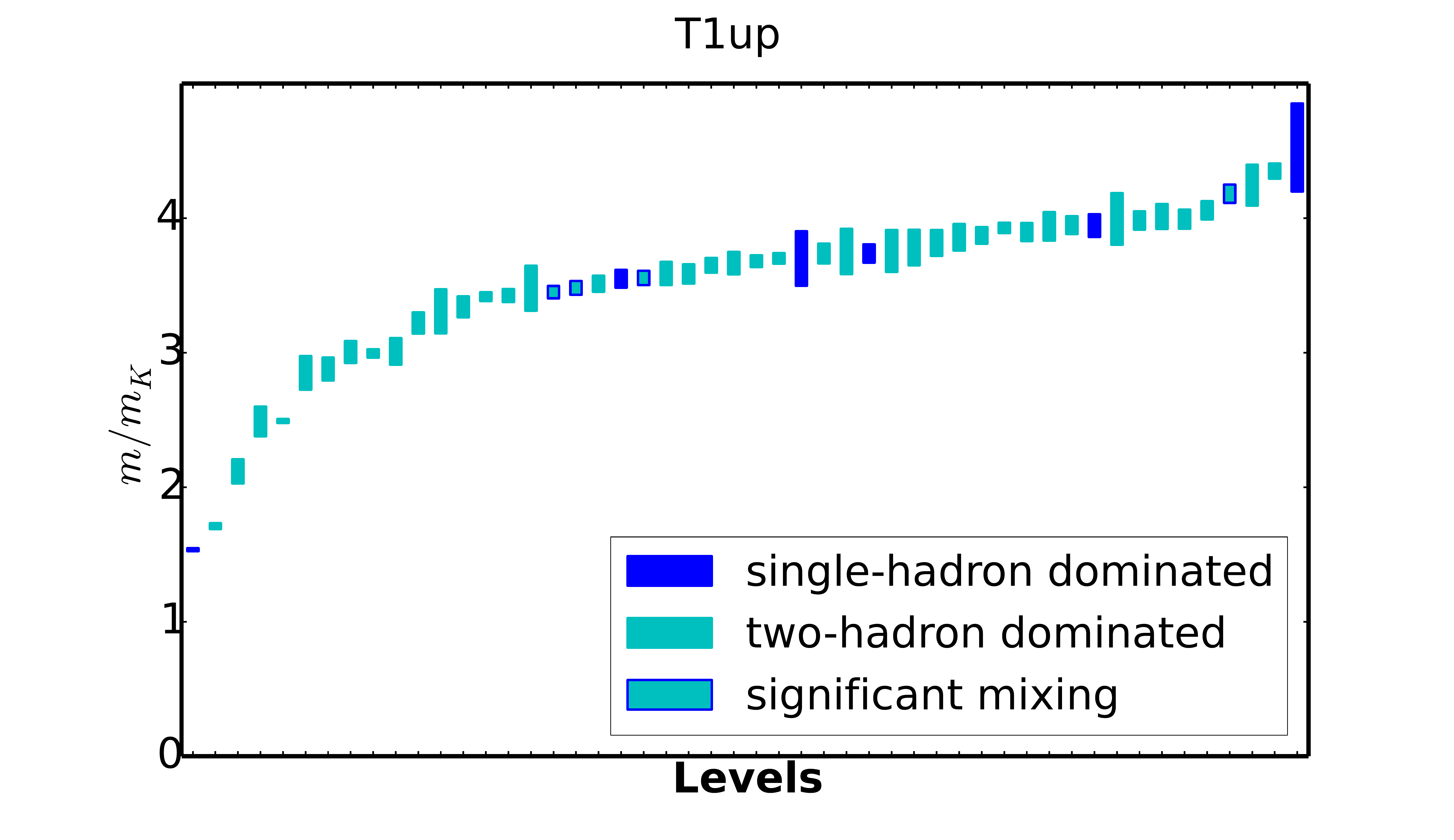} 
\caption{\label{fig:morningstar}  Left: The effective masses of the 16th excited to the 31st excited state in the $\rho$ channel obtained using a number of $\bar qq$ and meson-meson interpolators \cite{Morningstar:lat14}. Right: The energies of the   lowest 50 eigenstates in the same channel indicating their dominant Fock components.    }
\end{center}
\end{figure}

\underline{{\bf Light isovectors}}: An extensive study of various light isovector channels was performed with a number of  $\bar qq$ and  two-meson interpolators $\pi\pi$, $\eta\pi$, $\phi\pi$, $\bar KK$ in \cite{Morningstar:lat14}. The effective masses of 16th till 31st  eigenstate   in the $\rho$-meson channel are shown in Fig \ref{fig:morningstar} to demonstrate the high statistic quality of these high-lying states, obtained using the stochastic distillation. Energies of the lowest fifty eigenstates are  also  provided, together with their dominant Fock components based on $Z_i^n$.  The lowest level, for example, is related to the $\rho$-meson. 

\begin{figure}[htb] 
 \begin{center}
\includegraphics*[width=0.60\textwidth,clip]{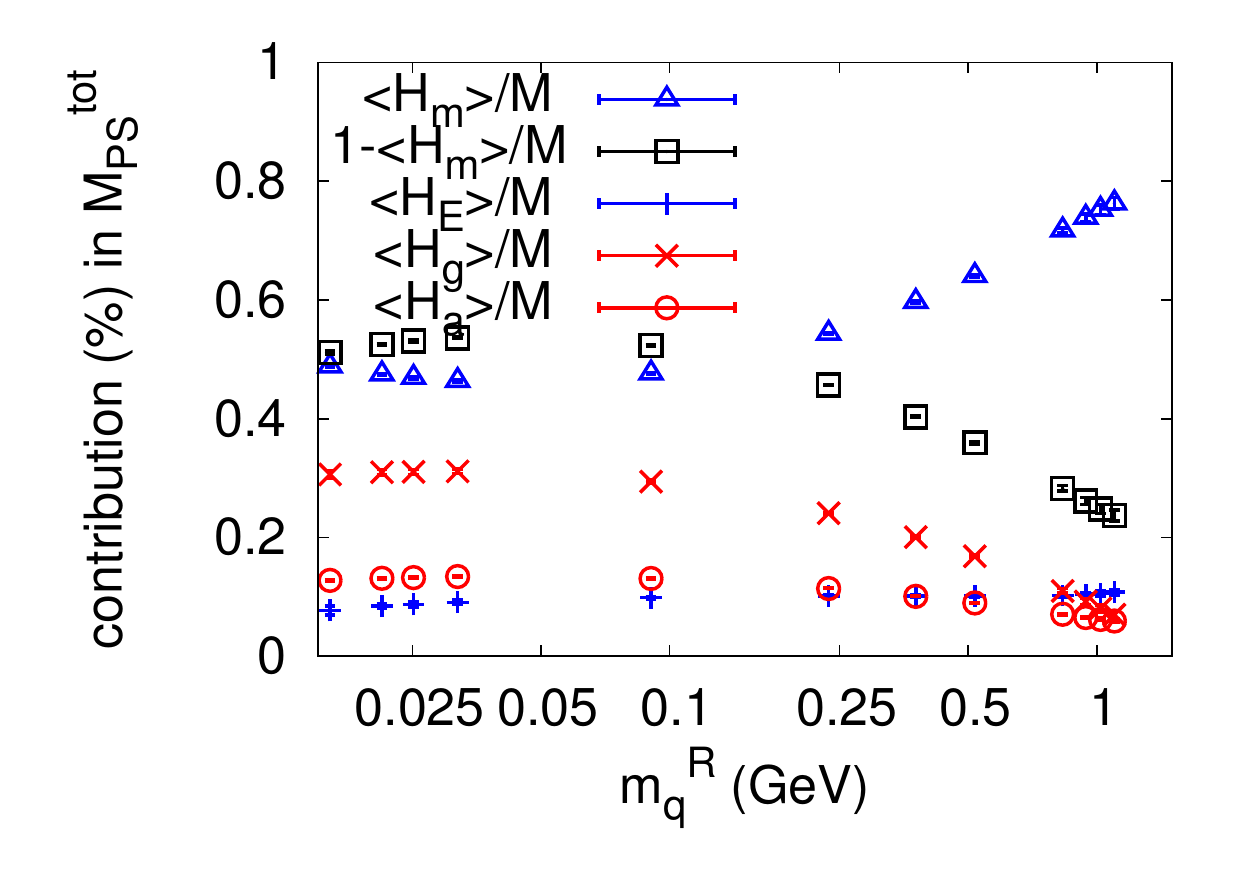}   
\caption{\label{fig:mass_decomposition}  The percentage of various contributions (defined in the main text) to the pseudoscalar meson mass as a function of the quark mass   \cite{Yang:2014xsa,Liu:lat14}.     }
\end{center}
\end{figure}

\section{Related topics}

\underline{{\bf Mass decomposition of mesons}}: The meson mass $M$ is decomposed to the quark mass part $H_m$, the  quark energy part $H_E$, the gluon field  energy part $H_g$ and    the QCD trace anomaly part $H_a$, i.e.  $M=-\langle T_{44}\rangle = \langle H_E\rangle+\langle H_m\rangle+ \langle H_g\rangle+  \langle H_a\rangle$ \cite{Yang:2014xsa,Liu:lat14}. Furthermore, these can be related via $\tfrac{1}{4}M=\tfrac{1}{4}  \langle H_m\rangle+  \langle H_a\rangle$. The $M$, $\langle H_E\rangle$ and $ \langle H_m\rangle$ are determined through the two- and three-point correlators directly on the lattice, while $\langle H_g\rangle$ and $ \langle H_a\rangle$ are obtained from the analytic relations presented in this paragraph. Proportions of each contribution to the pseudoscalar meson mass as a function of the quark mass is presented in Fig. \ref{fig:mass_decomposition}. For results on the vector meson and interesting physics conclusions, I refer to \cite{Yang:2014xsa,Liu:lat14}. \\

\underline{{\bf Extended QCD}}: The QCD is  reformulated as the so-called Extended-QCD by the help of additional bosonic fields, where the second theory is exactly equivalent to QCD \cite{Kaplan:2013dca}. It is shown that  the $N_c\to \infty$ limit  of the Extended-QCD respects two properties that were difficult to reconcile in the past:  (i) it makes direct connection to the constituent quark model with the constituent quark mass, (ii) it respects chiral symmetry rendering massless pion in the continuum limit.   The approach may prove valuable to understand why quark models often work well  even beyond the  $N_c\to \infty$ limit.

\section{Conclusions}
Spectra of hadrons below strong decay thresholds are calculated with unprecedented accuracy. Extensive results for multiplets within the single-hadron approach are available in light and heavy, meson and baryon sectors.  The first rigorous treatment of interesting near-threshold mesons $X(3872)$, $D_{s0}(2317)$ and $D_{s1}(2460)$ was performed during the past two years.  They appear as poles of the scattering matrix and their masses are close to the measured ones. Until recently, $\rho$ was the only resonance that was extracted using a Breit-Wigner type fit of the scattering phase shift, and it remains the only one that is mapped out in a great precision and by a number of collaborations. Strange, charmed and light-axial meson resonances  have been  rigorously treated for the first time recently.  The first simulation of the two-coupled channel  system was performed this year, where the poles corresponding to the strange mesons were extracted relying on the parametrization of the scattering matrix.  
Let me note that hadrons listed in this paragraph do not have manifestly exotic flavor and lattice QCD indeed presents  convincing support for them.

 On the other hand,   we have no reliable lattice evidence for flavor exotic states yet, in spite of the number of searches. These have thoroughly looked for  possible $\bar cc\bar du$ candidates with $J^{PC}=1^{+-}$ in the vicinity of $D\bar D^*$ threshold, where $Z_c^+(3900)$ was found experimentally. No $\bar cc\bar ss$ resonance was found in $J/\psi \phi$ scattering on the lattice, and no 
 $cc\bar d\bar u$ bound state in the $DD^*$ scattering. 
 
 However, there are many possibilities for improving searches of quarkonium-like states and other exotic states on the lattice. The available simulations present only the first steps in this direction.    The lattice community is welcome to  respond to the challenge set by the recent exciting experimental discoveries. This seems realistic for  the states relatively  near the strong decay thresholds, while much more challenging for higher lying states. 
 
\vspace{0.5cm}

 {\bf Acknowledgements}
 
 I would like to thank to Christine Davies, Carleton DeTar, Derek  Leinweber, Keh-Fei Liu, Daniel Mohler,  Colin Morningstar, Raul Briceno, Takeshi  Yamazaki, Andrea Guerrieri, Christopher Thomas and David Wilson  
 for valuable discussions and material related to this review talk. 
 I am grateful to  Christian B. Lang, Luka Leskovec, Daniel Mohler and Richard Woloshyn for the pleasure of collaborating on the described topics, and for reading this manuscript.  This work is supported by ARRS project number N1-0020 and FWF project number I1313-N27. 

\bibliographystyle{JHEP}
\bibliography{Lgt_lat14}


\end{document}